# TUNING TO $N = 2$ SUPERSYMMETRY IN THE SU(2) ADJOINT HIGGS-YUKAWA MODEL

**I. Montvay**[1]

Theoretical Physics Division, CERN
CH-1211 Geneva 23, Switzerland
and
Deutsches Elektronen-Synchrotron DESY,
Notkestr. 85, D-22603 Hamburg, Germany

## ABSTRACT

The $N = 2$ supersymmetric continuum limit is investigated in the SU(2) adjoint Higgs-Yukawa model using lattice perturbation theory. In the one-loop renormalization group equations a non-trivial infrared fixed point of coupling ratios is found. The phase structure at weak couplings is determined by a numerical study of the one-loop effective potential.

[1] e-mail address: montvay@surya20.cern.ch

# 1 Introduction

Renewed interest in exploiting the non-perturbative properties of gauge theories with extended supersymmetry has been triggered by recent work of Seiberg and Witten [1] (for earlier references see these papers, and for generalizations, see [2]). In order to understand the implications of these beautiful, exact results for lattice gauge theory, the lattice regularization of theories with extended supersymmetry has to be investigated. This can be done, as proposed some time ago by Curci and Veneziano [3], by letting the lattice spoil supersymmetry in the cut-off theory and requiring its restoration in the continuum limit. In the case of $N = 2$ and $N = 4$ extended supersymmetry the definition of the lattice theory is straightforward, because the difficult problem of reconciling the lattice with chiral symmetry does not arise due to the vector-like nature of these theories. (For a recent review of the chiral problem, see [4].)

The simplest prototype model with extended supersymmetry is the $N = 2$ supersymmetrized version of Yang-Mills theory with the SU(2) gauge group (SYM2). On the lattice this is embedded in a gauge model containing, besides the triplet gluinos, also scalar triplet fields. These latter have Yukawa couplings as well as quartic scalar couplings [5]; therefore this model belongs to the large class of Higgs-Yukawa models enjoying interest in the non-perturbative lattice literature (for a review see, for instance, [6]). The most general renormalizable quantum field theory containing SYM2 has seven couplings and three mass parameters. This makes the problem of parameter tuning for the continuum limit quite hard, even in this simplest example. Nevertheless, in the weak coupling region a very valuable guide is provided by the perturbative renormalization group equations (RGEs). Another useful tool in weak coupling theories with scalar fields is the perturbative effective potential, which gives information on the phase structure. The usefulness of lattice perturbation theory at weak couplings in lattice Higgs models is well known and widely explored (for a recent example, see [7]). In the present paper one-loop lattice perturbation theory will be used for the study of the SU(2) adjoint Higgs-Yukawa model in the vicinity of the supersymmetric fixed point.

Among the seven couplings of the renormalizable extension of SYM2, there are four quartic couplings. This extended set of quartic couplings plays a dual rôle: First, it is a complete set of couplings which is generated by quantum corrections from any smaller set of couplings once supersymmetry is broken. Second, the supersymmetric combination of couplings has flat directions which have to be cut-off in order that the path integral be well-defined.

An important condition for the existence of a supersymmetric continuum limit is the existence of a fixed point of coupling ratios corresponding to the supersymmetric relations. It is also necessary that this supersymmetric fixed point can be reached from at least one attractive direction in the ultraviolet. It has been shown in ref. [5] that both these conditions are fulfilled and there is exactly one attractive direction in the region with stable path integral. A closer investigation of the one-loop RGEs shows that the supersymmetric fixed point is accompanied by another fixed point of coupling ratios, which is actually an infrared fixed point. This will be shown in the next section of this paper.

The Feynman rules of lattice perturbation theory are defined in section 3. Section 4 is devoted to the derivation and numerical study of the one-loop lattice effective potential. In section 5 questions related to the restoration of global symmetries are investigated. The last section contains a discussion and the conclusions.

Throughout this paper the same notations will be used as in ref. [5] and some of the equations of this reference will only be referred to, without repeating them here.

## 2 Infrared fixed point of coupling ratios

The full set of one-loop RGEs for coupling ratios has been given in eq. (23) of ref. [5]. The coupling ratios are defined as

$$R_{A,B} \equiv \frac{G_{A,B}}{g} , \qquad r_{A,B} \equiv \frac{\lambda_{A,B}}{g^2} , \qquad r_{[AB]} \equiv \frac{\lambda_{[AB]}}{g^2} , \qquad r_{(AB)} \equiv \frac{\lambda_{(AB)}}{g^2} . \tag{1}$$

Here $g$ is the bare gauge coupling, $G_{A,B}$ are Yukawa couplings and $\lambda_{A,B,[AB],(AB)}$ quartic scalar couplings (for the lattice action see next section). The one-loop RGEs preserve the relations

$$R_A = R_B = 1 , \qquad r_A = r_B \equiv r_0 . \tag{2}$$

Furthermore, if these relations are fulfilled, we have

$$\begin{aligned} \frac{d}{dt}\left[4r_A r_B - (r_{[AB]} - r_{(AB)})^2\right] &= \frac{d}{dt}\left[4r_0^2 - (r_{[AB]} - r_{(AB)})^2\right] \\ = 704 r_0^3 - r_0\left[112(r_{[AB]} - r_{(AB)})^2 + 8 - 32 r_{(AB)}^2\right] &+ 4(r_{[AB]} - r_{(AB)}) \\ -32(r_{[AB]} - r_{(AB)})^3 - 16 r_{(AB)}^2 (r_{[AB]} - r_{(AB)}) &. \end{aligned} \tag{3}$$

The right-hand side here vanishes for

$$r_{[AB]} = r_{(AB)} + 2r_0 . \tag{4}$$

This implies that relation (4) is also preserved by the one-loop RGEs. It also gives a convergent path integral for positive $r_0$ and non-negative $r_{[AB]}$.

If the conditions in (2) and (4) are imposed, the one-loop RGEs are reduced to the simple form

$$\begin{aligned} \frac{dg^2}{dt} &= -\frac{8g^4}{16\pi^2} , \\ \frac{dr_0}{dt} &= \frac{g^2}{16\pi^2}\left(112 r_0^2 + 4 r_{(AB)}^2 + 16 r_0 r_{(AB)} - 1\right) , \\ \frac{dr_{(AB)}}{dt} &= \frac{g^2}{16\pi^2}\left(12 r_{(AB)}^2 + 96 r_0 r_{(AB)} - 3\right) . \end{aligned} \tag{5}$$

The solution of the first equation is

$$g^2(t) = \frac{g^2(t_0)}{1 + (t - t_0) g^2(t_0)/(2\pi^2)} . \tag{6}$$

The last two equations can be written in the variable $\tau \equiv \log(g_0^2/g^2)$ as

$$\frac{dr_0}{d\tau} = \frac{1}{8}\left(112r_0^2 + 4r_{(AB)}^2 + 16r_0 r_{(AB)} - 1\right) \ ,$$
$$\frac{dr_{(AB)}}{d\tau} = \frac{1}{8}\left(12r_{(AB)}^2 + 96r_0 r_{(AB)} - 3\right) \ . \tag{7}$$

The right-hand side of the equations in (7) vanish in the points

$$S \ : \qquad r_0 = 0 \ , \qquad r_{(AB)} = \frac{1}{2} \ ; \tag{8}$$

and

$$Q \ : \qquad r_0 = \frac{\sqrt{105}}{210} = 0.048795... \ , \qquad r_{(AB)} = \frac{\sqrt{105}}{30} = 0.341565... \ . \tag{9}$$

There are two other solutions obtained by reflections about the origin in the $(r_0, r_{(AB)})$-plane, but they do not correspond to convergent path integrals. The fixed point in (8) is the supersymmetric one in the special case of eqs. (2), (4). Note that the above construction implies that every fixed point of (7) supplemented by (2) and (4) is at the same time also a fixed point of the full set of RGEs in ref. [5].

The stability properties of the fixed points (8) and (9) can be deduced from the derivative matrix of the right-hand sides in (7):

$$\mathbf{D} = \begin{pmatrix} 28r_0 + 2r_{(AB)} & r_{(AB)} + 2r_0 \\ 12r_{(AB)} & 12r_0 + 3r_{(AB)} \end{pmatrix} \ . \tag{10}$$

The eigenvalues $\alpha_{1,2}$ and eigenvectors $\mathbf{e}_{1,2}$ of this matrix in the supersymmetric point (8) are the following:

$$\alpha_1 = -1/2 \quad , \quad \mathbf{e}_1 = (1, -3) \ ,$$
$$\alpha_2 = 3 \quad , \quad \mathbf{e}_2 = (1, 4) \ . \tag{11}$$

Since, according to eq. (6), for $t \to \infty$ also $\tau \to \infty$, in the continuum limit the negative eigenvalues are attractive and the positive ones repulsive. The direction $\mathbf{e}_1$ corresponds to the attractive direction in the supersymmetric fixed point found in ref. [5]. In the other fixed point, given by (9), we get

$$\alpha_1 = (5\sqrt{105} + 3\sqrt{161})/28 = 3.1893... \quad , \quad \mathbf{e}_1 = (1, -1/2 + \sqrt{345}/6 = 2.5957...) \ ,$$

$$\alpha_2 = (5\sqrt{105} - 3\sqrt{161})/28 = 0.4703... \quad , \quad \mathbf{e}_2 = (1, -1/2 - \sqrt{345}/6 = -3.5957...) \ . \tag{12}$$

These are both repulsive in the ultraviolet and hence attractive in the infrared. The existence of this infrared fixed point in the RGEs of coupling ratios has interesting consequences for the supersymmetric continuum limit. These will be discussed in section 6 in connection with the phase structure at weak couplings.

The two fixed points $S$ in (8) and $Q$ in (9) determine the flow pattern of renormalization group trajectories in the physically interesting range $r_0 > 0$, $r_{(AB)} \geq -2r_0$. For $r_0, r_{(AB)} > 0$ this is illustrated by numerical solutions of eq. (7) in fig. 1. The figure nicely shows the separating line of the flow which connects the infrared fixed point $Q$ with the supersymmetric fixed point $S$. This line leads to the continuum limit with supersymmetric coupling ratios. It approaches $S$ from the ultraviolet attractive direction $\mathbf{e}_1$ in (11). The flow line approaching $S$ from the infrared attractive direction $\mathbf{e}_2$ is also shown. It separates the flow lines going in the infrared to $Q$ from those going to the negative $r_0$ region.

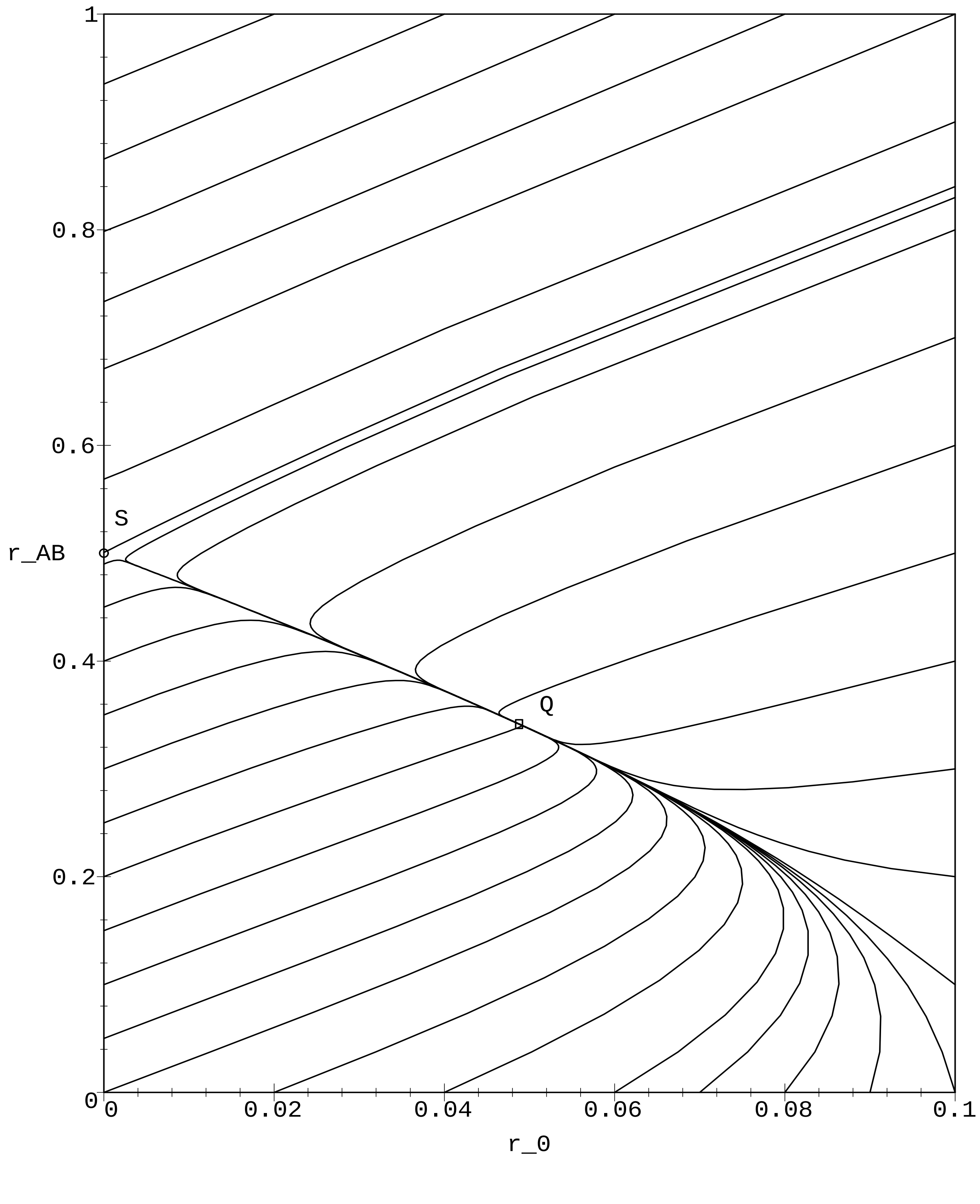

Figure 1: The renormalization group trajectories in the $(r_0, r_{(AB)})$-plane around the two fixed points $S$ and $Q$. The latter is attractive in the infrared limit. This determines the flow directions.

# 3 Lattice perturbation theory

The physical fields in SYM2 are: $A^r_{x\mu}$, $\psi^r_x$, $\overline{\psi}^r_x$, $A^r_x$, $B^r_x$. The gauge field $A^r_{x\mu}$, $\mu \in \{1,2,3,4\}$, $r \in \{1,2,3\}$ is represented on the lattice by the SU(2) matrix on links $U_{x\mu} \equiv \exp(igT_r A^r_{x\mu})$, with the SU(2) generators $T_r = \tau_r/2$. The lattice spacing (usually denoted by $a$) is set to 1 throughout this paper; in other words every dimensional quantity is measured in lattice units. The fermion fields can be represented by the four-component Dirac field in the adjoint (triplet) representation $\psi^r_x$, $\overline{\psi}^s_x$, $r,s \in \{1,2,3\}$. The real scalar and pseudoscalar triplet fields are denoted by $A^r_x$, $r \in \{1,2,3\}$ and $B^s_x$, $s \in \{1,2,3\}$, respectively. Given this set of fields, we shall consider the most general renormalizable quantum field theory, which respects gauge invariance and certain global symmetries, as e. g. parity conservation. This is the SU(2) Higgs-Yukawa model studied in ref. [5].

The lattice action has been given in eqs. (1)-(4) of ref. [5]. It can be written as

$$S = S_g + S_f + S_s \ . \tag{13}$$

The standard Wilson action for the gauge field $S_g$ is a sum over the plaquettes

$$S_g = \beta \sum_{pl} \left( 1 - \frac{1}{2} \mathrm{Tr}\, U_{pl} \right) \ , \tag{14}$$

with the bare gauge coupling given by $\beta \equiv 4/g^2$. In a triplet notation and with the usual perturbative field normalization, the fermionic part $S_f$ is

$$S_f = \sum_x \Big\{ (m_\psi + 4r)\overline{\psi}^r_x \psi^r_x + i\epsilon_{rst}\overline{\psi}^r_x (G_A A^s_x + iG_B \gamma_5 B^s_x)\psi^t_x$$

$$-\frac{1}{2}\sum_{\mu=1}^{4} \left[ \overline{\psi}^r_{x+\hat{\mu}} V_{rs,x\mu}(r+\gamma_\mu)\psi^s_x + \overline{\psi}^r_x V^{-1}_{rs,x\mu}(r-\gamma_\mu)\psi^s_{x+\hat{\mu}} \right] \Big\} \ . \tag{15}$$

Here $m_\psi$ is the bare fermion (Dirac) mass, $0 < r \leq 1$ is the irrelevant Wilson parameter removing the fermion doublers in the continuum limit, and the O(3) matrix for the gauge-field link is defined as

$$V_{rs,x\mu} \equiv 2\mathrm{Tr}(U^\dagger_{x\mu} T_r U_{x\mu} T_s) = V^*_{rs,x\mu} = V^{-1T}_{rs,x\mu} \ . \tag{16}$$

The scalar part of the lattice action $S_s$ is

$$S_s = \sum_x \Big\{ \frac{1}{2}(m_A^2 + 8)A^r_x A^r_x + \frac{1}{2}(m_B^2 + 8)B^r_x B^r_x - \sum_{\mu=1}^{4} \left[ A^r_{x+\hat{\mu}} V_{rs,x\mu} A^s_x + B^r_{x+\hat{\mu}} V_{rs,x\mu} B^s_x \right]$$

$$+\lambda_A (A^r_x A^r_x)^2 + \lambda_B (B^r_x B^r_x)^2 + \lambda_{[AB]} A^r_x A^r_x B^s_x B^s_x - \lambda_{(AB)} (A^r_x B^r_x)^2 \Big\} \ . \tag{17}$$

Here $m_A$ and $m_B$ denote the bare mass parameters for the two scalar fields, respectively.

In the symmetric phase with vanishing vacuum expectation values the propagators for the scalar, fermion and gauge fields are, respectively,

$$\Delta^{A,B}_{x_1,x_2} = \frac{1}{\Omega} \sum_k e^{ik\cdot(x_1-x_2)} \left( m^2_{A,B} + \hat{k}^2 \right) \ ,$$

$$\Delta^{\psi}_{x_1,x_2} = \frac{1}{\Omega} \sum_k e^{ik\cdot(x_1-x_2)} \frac{m_\psi + \frac{r}{2}\hat{k}^2 - i\gamma\cdot\bar{k}}{(m_\psi + \frac{r}{2}\hat{k}^2)^2 + \bar{k}^2} = \frac{1}{\Omega} \sum_k e^{ik\cdot(x_1-x_2)} \left( m_\psi + \frac{r}{2}\hat{k}^2 + i\gamma\cdot\bar{k} \right)^{-1} \ ,$$

$$\Delta^{g}_{x_1\mu_1,x_2\mu_2} = \frac{1}{\Omega} \sum_k e^{ik\cdot(x_1-x_2)+i(k_{\mu_1}-k_{\mu_2})/2} \frac{1}{\hat{k}^2} \left[ \delta_{\mu_1\mu_2} - (1-\alpha)\frac{\hat{k}_{\mu_1}\hat{k}_{\mu_2}}{\hat{k}^2} \right] \ , \tag{18}$$

where, as usual, $\Omega$ is the number of lattice points, $\alpha$ is the gauge-fixing parameter and

$$\hat{k}_\mu \equiv 2\sin(\frac{k_\mu}{2}) \ , \qquad \bar{k}_\mu \equiv \sin(k_\mu) \ . \tag{19}$$

The vertices can be read from the action (13)–(15) after expanding the gauge links, with $|A_{x\mu}|^2 \equiv A^r_{x\mu} A^r_{x\mu}$, according to

$$U_{x\mu} = \cos\left(\frac{g}{2}|A_{x\mu}|\right) + 2iT_r \frac{A^r_{x\mu}}{|A_{x\mu}|} \sin\left(\frac{g}{2}|A_{x\mu}|\right) \equiv a^0_{x\mu} + 2iT_r a^r_{x\mu} \ ;$$

$$V_{rs,x\mu} = \delta_{rs}\left(a^0_{x\mu}a^0_{x\mu} - a^t_{x\mu}a^t_{x\mu}\right) + 2\epsilon_{rst} a^0_{x\mu} a^t_{x\mu} + 2a^r_{x\mu}a^s_{x\mu}$$

$$= \delta_{rs} + g\epsilon_{rst} A^t_{x\mu} + \frac{g^2}{2}\left(A^r_{x\mu}A^s_{x\mu} - \delta_{rs}A^t_{x\mu}A^t_{x\mu}\right) + \ldots \ . \tag{20}$$

The Feynman rules in the phases with non-vanishing vacuum expectation values

$$v_A \delta_{3r} = \langle A^r_x \rangle \neq 0 \ , \qquad v_B \delta_{3r} = \langle B^r_x \rangle \neq 0 \tag{21}$$

are, of course, somewhat more complicated. As has been discussed in ref. [5], according to the classical potential there are three different phases with the symmetry-breaking pattern SU(2)$\to$U(1): phase $A$ with $v_A \neq 0$, $v_B = 0$, phase $B$ with $v_B \neq 0$, $v_A = 0$ and phase $AB$ with $v_A \neq 0$, $v_B \neq 0$ and parallel vacuum expectation values as in (21). (The phase with orthogonal vacuum expectation values and complete breaking of the SU(2) gauge symmetry is not relevant to supersymmetry.) It can be expected that for weak couplings the phase structure remains qualitatively the same. In fact, this is the main question which will be investigated in the present paper within the framework of lattice perturbation theory.

In order to illustrate the Feynman rules in these phases with Higgs mechanism, let us briefly consider the most interesting case of phase $AB$. The other cases are simpler and can be treated similarly. Here only a few remarks will be included about them, which are relevant in some important, degenerate situations. Let us split the scalar field according to

$$A^r_x \equiv \bar{A}^r_x + v_A \delta_{3r} \ , \qquad B^r_x \equiv \bar{B}^r_x + v_B \delta_{3r} \tag{22}$$

and consider the gauge boson propagator. The gauge-fixing function is conveniently introduced in such a way that the scalar-gauge mixing is cancelled [8]. This is achieved by taking, for the gauge-fixing part of the action,

$$S_{gf} = \frac{1}{2\alpha} \sum_x f_{rx} f_{rx} \tag{23}$$

with the gauge-fixing function

$$f_{rx} \equiv \sum_{\mu=1}^{4} (A^r_{x\mu} - A^r_{x-\hat{\mu},\mu}) + \alpha g \epsilon_{r3t} (v_A \bar{A}^t_x + v_B \bar{B}^t_x) \ . \tag{24}$$

With this choice the gauge boson propagator becomes

$$\Delta^g_{x_1\mu_1,x_2\mu_2} = \frac{1}{\Omega} \sum_k e^{ik\cdot(x_1-x_2)+i(k_{\mu_1}-k_{\mu_2})/2} \frac{1}{m^2_{g,r} + \hat{k}^2} \left[ \delta_{\mu_1\mu_2} - (1-\alpha) \frac{\hat{k}_{\mu_1}\hat{k}_{\mu_2}}{\alpha m^2_{g,r} + \hat{k}^2} \right] \ . \tag{25}$$

Here the gauge boson masses $m_{g,r}$ are given by

$$m^2_{g,1} = m^2_{g,2} = g^2(v^2_A + v^2_B) \ , \qquad m^2_{g,3} = 0 \ . \tag{26}$$

The mass square matrix for the scalar fields $(A, B)$ can be obtained by introducing the shifted fields $(\bar{A}, \bar{B})$ defined in (22) into the lattice action and collecting the quadratic terms in the minimum of the potential. Adding also the contribution of the gauge-fixing term proportional to $\alpha$, one obtains for the mass square matrix in the subspace with isospin indices 1,2 containing the Goldstone bosons:

$$M^2_{(12)} = \begin{pmatrix} 2\lambda_{(AB)} v^2_B + \alpha g^2 v^2_A & -2\lambda_{(AB)} v_A v_B + \alpha g^2 v_A v_B \\ -2\lambda_{(AB)} v_A v_B + \alpha g^2 v_A v_B & 2\lambda_{(AB)} v^2_A + \alpha g^2 v^2_B \end{pmatrix} \ . \tag{27}$$

In the orthogonal subspace with isospin index 3 we get

$$M^2_{(3)} = \begin{pmatrix} 8\lambda_A v^2_A & 4(\lambda_{[AB]} - \lambda_{(AB)}) v_A v_B \\ 4(\lambda_{[AB]} - \lambda_{(AB)}) v_A v_B & 8\lambda_B v^2_B \end{pmatrix} \ . \tag{28}$$

In the Landau gauge ($\alpha = 0$), these give for the momentum space scalar propagator

$$\tilde{G}^{AB} \equiv (M^2_{(123)} + \hat{k}^2)^{-1} \tag{29}$$

the non-vanishing matrix elements

$$\tilde{G}^{AB}_{11} = \tilde{G}^{AB}_{22} = \frac{\hat{k}^2 + 2\lambda_{(AB)} v^2_A}{\hat{k}^2[\hat{k}^2 + 2\lambda_{(AB)}(v^2_A + v^2_B)]} \ ,$$

$$\tilde{G}^{AB}_{44} = \tilde{G}^{AB}_{55} = \frac{\hat{k}^2 + 2\lambda_{(AB)} v^2_B}{\hat{k}^2[\hat{k}^2 + 2\lambda_{(AB)}(v^2_A + v^2_B)]} \ ,$$

$$\tilde{G}^{AB}_{14} = \tilde{G}^{AB}_{25} = \tilde{G}^{AB}_{41} = \tilde{G}^{AB}_{52} = \frac{2\lambda_{(AB)} v_A v_B}{\hat{k}^2[\hat{k}^2 + 2\lambda_{(AB)}(v^2_A + v^2_B)]} \ ,$$

$$\tilde{G}^{AB}_{33} = \frac{\hat{k}^2 + 8\lambda_B v_B^2}{\hat{k}^4 + 8\hat{k}^2(\lambda_A v_A^2 + \lambda_B v_B^2) + 16 v_A^2 v_B^2 [4\lambda_A\lambda_B - (\lambda_{[AB]} - \lambda_{(AB)})^2]} ,$$

$$\tilde{G}^{AB}_{66} = \frac{\hat{k}^2 + 8\lambda_A v_A^2}{\hat{k}^4 + 8\hat{k}^2(\lambda_A v_A^2 + \lambda_B v_B^2) + 16 v_A^2 v_B^2 [4\lambda_A\lambda_B - (\lambda_{[AB]} - \lambda_{(AB)})^2]} ,$$

$$\tilde{G}^{AB}_{36} = \tilde{G}^{AB}_{63} = \frac{-4(\lambda_{[AB]} - \lambda_{(AB)}) v_A v_B}{\hat{k}^4 + 8\hat{k}^2(\lambda_A v_A^2 + \lambda_B v_B^2) + 16 v_A^2 v_B^2 [4\lambda_A\lambda_B - (\lambda_{[AB]} - \lambda_{(AB)})^2]} . \qquad (30)$$

The indices are defined here in such a way that the first three belong to $A$ and the last three to $B$. The notation $\hat{k}^{2n}$ means, of course, $(\hat{k}^2)^n$.

An important, special, degenerate case of the phase structure occurs when the $AB$-phase collapses to a line in the $(m_A^2, m_B^2)$-space and, therefore, the $A$- and $B$-phases touch each other. At tree level this happens when

$$4\lambda_A\lambda_B = (\lambda_{[AB]} - \lambda_{(AB)})^2 , \qquad m_A^2\sqrt{\lambda_B} = m_B^2\sqrt{\lambda_A} . \qquad (31)$$

In this case, on the line unifying all three phases, there is a *minimum valley* where the vacuum expectation values of the $A$- and $B$-fields are on an ellipse

$$v_A^2\sqrt{\lambda_A} + v_B^2\sqrt{\lambda_B} = -\frac{m_A^2}{4\sqrt{\lambda_A}} = -\frac{m_B^2}{4\sqrt{\lambda_B}} . \qquad (32)$$

In fact, the perturbation theory in this degenerate situation can be considered as a special case of the perturbation theories in any of the three phases. For definiteness, one can transform the vacuum expectation value to the $A$-direction by an axial $\mathrm{U}(1)_A$ transformation

$$\psi'_x = e^{-i\phi\gamma_5}\psi_x , \qquad \overline{\psi}'_x = \overline{\psi}_x e^{-i\phi\gamma_5} ,$$

$$A'_x = \cos(2\phi)A_x - \sin(2\phi)B_x , \qquad B'_x = \sin(2\phi)A_x + \cos(2\phi)B_x . \qquad (33)$$

Then, for instance, (30) becomes

$$\tilde{G}^{AB}_{11} = \tilde{G}^{AB}_{22} = \tilde{G}^{AB}_{66} = \frac{1}{\hat{k}^2} , \qquad \tilde{G}^{AB}_{33} = \frac{1}{\hat{k}^2 + 8\lambda_A v_A^2} , \qquad \tilde{G}^{AB}_{44} = \tilde{G}^{AB}_{55} = \frac{1}{\hat{k}^2 + 2\lambda_{(AB)} v_A^2} . \qquad (34)$$

As one can see, besides the Goldstone bosons $\bar{A}^1, \bar{A}^2$ there is also a third zero-mass boson $\bar{B}^3$. This is one of the scalar $N = 2$ superpartners of the massless photon field $A^3_\mu$. Since it is associated with the global axial $\mathrm{U}(1)_A$ transformation, we can call it *photo-axion.* The other scalar superpartner of the photon field ($\bar{A}^3$) is an excitation orthogonal to the minimum valley in the plane of the two vacuum expectation values. Since it has to do with the length of the vacuum expectation values, we can call it *photo-dilaton.* In the same spirit, the scalar members of an $N = 2$ vector (gauge-) supermultiplet can be called *gauge-scalars*, *gluo-scalars* or *photo-scalars.* In perturbation theory the photo-dilaton is not exactly massless, but as can be seen from (34), its mass square is proportional to $\lambda_A$. Near the supersymmetric continuum limit $g^2 \to 0$, according to eqs. (7)–(11), $\lambda_A$ goes to zero as $\lambda_A \propto g$. Of course, these are just masses in the propagators of perturbation theory and the real question is how the physical masses of the photo-axion and photo-dilaton behave in the continuum limit near the supersymmetric fixed point. For answering this question, one has to first see whether the appropriate phase structure is reproduced at all near the continuum limit (see next section).

Returning to the Feynman rules, in the $AB$-phase the inverse propagator of the fermion field in momentum space is

$$\tilde{G}^{\psi-1}_{r_1 r_2} = \delta_{r_1 r_2} \left( m_\psi + \frac{r}{2}\hat{k}^2 + i\gamma \cdot \bar{k} \right) + i\epsilon_{r_1 3 r_2}(G_A v_A + i\gamma_5 G_B v_B) \ . \tag{35}$$

After introducing the notations

$$\tilde{G}^\psi_1 = \frac{m_{\psi k} + G_A v_A - i\gamma_5 G_B v_B - i\gamma \cdot \bar{k}}{(m_{\psi k} + G_A v_A)^2 + G_B^2 v_B^2 + \bar{k}^2} \ , \qquad \tilde{G}^\psi_2 = \frac{m_{\psi k} - G_A v_A + i\gamma_5 G_B v_B - i\gamma \cdot \bar{k}}{(m_{\psi k} - G_A v_A)^2 + G_B^2 v_B^2 + \bar{k}^2} \ ,$$

$$\tilde{G}^\psi_3 = \frac{m_{\psi k} - i\gamma \cdot \bar{k}}{m_{\psi k}^2 + \bar{k}^2} \ , \qquad m_{\psi k} = m_\psi + \frac{r}{2}\hat{k}^2 \ , \tag{36}$$

the fermion propagator in momentum space can be written as

$$\tilde{G}^\psi = \begin{pmatrix} \frac{1}{2}(\tilde{G}^\psi_1 + \tilde{G}^\psi_2) & \frac{i}{2}(\tilde{G}^\psi_1 - \tilde{G}^\psi_2) & 0 \\ -\frac{i}{2}(\tilde{G}^\psi_1 - \tilde{G}^\psi_2) & \frac{1}{2}(\tilde{G}^\psi_1 + \tilde{G}^\psi_2) & 0 \\ 0 & 0 & \tilde{G}^\psi_3 \end{pmatrix} \ . \tag{37}$$

Besides the propagators, one also has to determine the vertices. They can be obtained also in the Higgs phases by simply expanding the lattice action in powers of the fields $A^r_{x\mu}$ in (20), $\bar{A}^r_x, \bar{B}^r_x$ in (22), and $\psi^r_x, \overline{\psi}^r_x$.

# 4 Effective potential and phase structure

The existence of a supersymmetric continuum limit in a renormalizable lattice theory, with the same set of fields and a set of couplings and mass parameters broader than the supersymmetric *target theory*, imposes several conditions on the behaviour of the bare theory for small lattice spacings. In the case of an asymptotically free continuum limit, as required for SYM2, the continuum limit is defined by approaching the free *Gaussian fixed point* at zero couplings. The first condition to be satisfied is the existence of a fixed point in the RGEs for coupling ratios corresponding to the supersymmetry relation between the gauge coupling ($g$) and the other relevant couplings. This fixed point should occur for zero gauge coupling and should have at least one attractive direction in the $g \to 0$ limit, in order that it can be reached from non-zero $g$. As has been shown in ref. [5], this condition is satisfied in SYM2 (see also section 2 of this paper).

The second condition for the existence of the supersymmetric continuum limit is that the appropriate phase structure, which arises in the classical theory from the classical potential, is also reproduced near the Gaussian fixed point in the quantum theory. Besides the trivial requirement of second-order nature of the phase transitions, which makes the continuum limit possible in general, there are also more subtle conditions to satisfy. These follow from the necessity of reproducing the *flat directions* in the effective potential corresponding to the *quantum moduli space* (QMS) studied in ref. [1]. The *classical moduli space* (CMS) of SYM2 arises as a particular limit of the minima of the classical potential in the SU(2) Higgs-Yukawa model. The compact flat direction belonging to the angle of the vacuum expectation values of the scalar $A$- and $B$-fields corresponds to the minimum valley, which appears if the phases $A$, $B$ and $AB$ become degenerate on a line in the plane of bare scalar mass squares $(m_A^2, m_B^2)$ (or scalar hopping parameters $(\kappa_A, \kappa_B)$). The non-compact flat direction belonging to the overall length of the vacuum expectation values is obtained in the limit where the quartic scalar couplings stabilizing the path integral $(\lambda_A, \lambda_B)$ tend to zero. As already discussed in the previous section, the existence of these two flat directions in the QMS requires that the physical mass of the photo-axion and photo-dilaton be zero in the continuum limit. In summary: the requirement on the phase structure is the existence of second-order phase transitions having the same topology as in the classical theory, in particular, showing the degeneracy of the three phases $A$, $B$ and $AB$ in a line of the $(m_A^2, m_B^2)$-plane.

The phase structure of lattice quantum field theories can be effectively exploited by numerical simulations. Another useful tool, which is reliable in models with scalar fields with weak couplings, is the study of the effective potential in lattice perturbation theory. Previous experience shows that the nature and position of phase transitions in bare parameter space is quite well reproduced by the one-loop approximation. For instance, in the SU(2) fundamental Higgs model in the physically relevant region $\beta \simeq 8$ - $10$ and $\lambda \simeq 10^{-4}$ - $10^{-3}$ both the invariant effective potential (in the unitary gauge) and the Landau gauge ($\alpha = 0$) effective potential work well [7]. (In fact, in the published version of ref. [7] only the invariant effective potential is considered, but very similar results can be obtained in the Landau gauge too.) In the present context the study of lattice perturbation theory is even more important than in simple Higgs models. This is because the numerical simulation is much more expensive, due to the presence of fermions and the large number of parameters. In the present section I shall consider the one-loop lattice effective potential in the Landau gauge.

The one-loop effective action for the bosonic fields is given in general by

$$\Gamma_{1-\mathrm{loop}}[A^r_{x\mu}, A^r_x, B^r_x] = S_b[A^r_{x\mu}, A^r_x, B^r_x] + \frac{1}{2}\mathrm{Tr}\log\{D_b[A^r_{x\mu}, A^r_x, B^r_x]\Delta^b\}$$
$$-\mathrm{Tr}\log\{M^{FP}[A^r_{x\mu}, A^r_x, B^r_x]\Delta^{FP}\} - \mathrm{Tr}\log\{M^{\psi}[A^r_{x\mu}, A^r_x, B^r_x]\Delta^{\psi}\} \; . \tag{38}$$

Here $S_b \equiv S_g + S_s$ is the bosonic part of the lattice action in eqs. (13)–(17), $D_b$ its second derivative matrix, $\Delta^b$ the bosonic propagator matrix, $M^{FP}$ the Fadeev-Popov matrix for the ghost fields with the ghost propagator matrix $\Delta^{FP}$, $\Delta^{\psi}$ the fermion propagator matrix and $M^{\psi}$ the fermion matrix in the action defined by writing the fermionic part as

$$S_f \equiv \sum_{rx,sy} \overline{\psi}^s_y M^{\psi}_{sy,rx} \psi^r_x \; . \tag{39}$$

The one-loop effective potential is defined for $x$-independent scalar fields $A^r, B^r$ by

$$V^{1-\mathrm{loop}}_{\mathrm{eff}} \equiv \frac{1}{\Omega}\Gamma_{1-\mathrm{loop}}[A^r_{x\mu} = 0, A^r_x = A^r, B^r_x = B^r] \; . \tag{40}$$

Up to now these are the unrenormalized expressions. In the vicinity of second-order phase transitions, where the masses and vacuum expectation values tend to zero in lattice units, the bare perturbation theory in not convergent. This is due to the large logarithms $\log(am)$, where $am$ is a typical mass or vacuum expectation value in lattice units. Here we would like to determine the phase structure in bare parameter space; therefore, we shall not completely switch to renormalized perturbation theory, but only replace the masses in the propagators by the renormalized ones. This can be achieved in the loop expansion by adding to the action appropriate external sources proportional to the mass terms, which are quadratic in the bosonic fields. It is a general experience in lattice quantum field theory that the location and nature of phase transitions in bare parameter space are then well reproduced. In fact, an even simpler approximation is working well, when the masses in lattice units are set to zero everywhere in the propagators. This is the approximation I shall exploit in what follows. The effective potential will be calculated in the symmetric phase without scalar field expectation values and will be studied numerically in the vicinity of the phase transition hypersurfaces.

The technical advantage of considering the Landau gauge, with gauge parameter $\alpha = 0$, is that there is no mixed scalar-gauge loop contribution and the Fadeev-Popov ghost loop contribution vanishes, as well. Hence the one-loop effective potential in the Landau gauge can be written as the sum of scalar, fermion and gauge loops:

$$V_{\rm eff}^{\rm 1-loop} = V_{\rm eff}^{\rm scalar} + V_{\rm eff}^{\rm fermion} + V_{\rm eff}^{\rm gauge} \ . \qquad (41)$$

Let us first consider the scalar contribution, which is in fact the most complicated one out of the three. According to eqs. (38)–(40) it can be written as

$$V_{\rm eff}^{\rm scalar} = \frac{1}{2\Omega}\sum_k \log\det\left\{ \mathbf{1} + \begin{pmatrix} V_{AA} & V_{AB} \\ V_{BA} & V_{BB} \end{pmatrix} \begin{pmatrix} (m_A^2+\hat{k}^2)^{-1} & 0 \\ 0 & (m_B^2+\hat{k}^2)^{-1} \end{pmatrix} \right\}$$
$$\equiv \frac{1}{2\Omega}\sum_k \log \mathcal{D}_s \ . \qquad (42)$$

It follows from the lattice action in eq. (17) that the $3\otimes 3$ coupling matrices $V_{..}$ are given by

$$V_{AA,r_1r_2} = \delta_{r_1r_2}\Big(4\lambda_A(AA) + 2\lambda_{[AB]}(BB)\Big) + 8\lambda_A A^{r_1}A^{r_2} - 2\lambda_{(AB)}B^{r_1}B^{r_2} \ ,$$
$$V_{BB,r_1r_2} = \delta_{r_1r_2}\Big(4\lambda_B(BB) + 2\lambda_{[AB]}(AA)\Big) + 8\lambda_B B^{r_1}B^{r_2} - 2\lambda_{(AB)}A^{r_1}A^{r_2} \ ,$$
$$V_{AB,r_1r_2} = 4\lambda_{[AB]}A^{r_1}B^{r_2} - 2\lambda_{(AB)}\delta_{r_1r_2}(AB) - 2\lambda_{(AB)}B^{r_1}A^{r_2} \ ,$$
$$V_{BA,r_1r_2} = 4\lambda_{[AB]}B^{r_1}A^{r_2} - 2\lambda_{(AB)}\delta_{r_1r_2}(AB) - 2\lambda_{(AB)}A^{r_1}B^{r_2} \ . \qquad (43)$$

Here for the scalar products of isospin vectors the notation $(AB) \equiv A^sB^s$ etc. has been introduced. The evaluation of the determinant $\mathcal{D}_s$ in eq. (42) is possible by an algebraic manipulation program, but the resulting expression is rather long. In the important special case $\lambda_B = \lambda_A$, $\lambda_{[AB]} = \lambda_{(AB)}+2\lambda_A$, which corresponds to the relations (2), (4) for the coupling ratios considered in section 2, the result is given in the appendix. Further specializing to

$$\lambda_B = \lambda_A = 0 \ , \qquad \lambda_{[AB]} = \lambda_{(AB)} = \frac{g^2}{2} \ , \qquad (44)$$

we get

$$\mathcal{D}_s = 1 + 2g^2 D_1 + g^4(D_1^2 - 3D_2) - 4g^6 D_1 D_2 - g^8(D_1^2 + D_2)D_2 + 2g^{10} D_1 D_2^2 + 3g^{12} D_2^3 \ . \quad (45)$$

Here the abbreviations

$$D_1 \equiv D_1(A,B,k) \equiv \frac{(AA)}{m_B^2 + \hat{k}^2} + \frac{(BB)}{m_A^2 + \hat{k}^2} \ ,$$

$$D_2 \equiv D_2(A,B,k) \equiv \frac{(AA)(BB) - (AB)^2}{(m_A^2 + \hat{k}^2)(m_B^2 + \hat{k}^2)} \quad (46)$$

have been used.

The fermion-loop contribution in eq. (41) can be similarly calculated to give

$$V_{\rm eff}^{\rm fermion} = -\frac{2}{\Omega} \sum_k \log \mathcal{D}_f \ , \quad (47)$$

where

$$\mathcal{D}_f = 1 + D_\psi^2 \left[ 2G_A^2(AA)(\bar{k}^2 - m_{\psi k}^2) + 2G_B^2(BB)(\bar{k}^2 + m_{\psi k}^2) \right]$$
$$+ D_\psi^4 \left[ G_A^4(AA)^2(\bar{k}^2 + m_{\psi k}^2)^2 + G_B^4(BB)^2(\bar{k}^2 + m_{\psi k}^2)^2 \right.$$
$$\left. + 2G_A^2 G_B^2(AA)(BB)(\bar{k}^4 - m_{\psi k}^4) + 4G_A^2 G_B^2 (AB)^2(\bar{k}^2 + m_{\psi k}^2) m_{\psi k}^2 \right] \quad (48)$$

and with (36) we defined

$$D_\psi \equiv (m_{\psi k}^2 + \bar{k}^2)^{-1} \ . \quad (49)$$

Finally, the contribution of the gauge-boson loop is given by

$$V_{\rm eff}^{\rm gauge} = \frac{3}{2\Omega} \sum_k \log \mathcal{D}_g \ , \quad (50)$$

where

$$\mathcal{D}_g = 1 + 2g^2 D'_1 + g^4(D'^2_1 + D'_2) + g^6 D'_1 D'_2 \ . \quad (51)$$

Here the notations are:

$$D'_1 \equiv D'_1(A,B,k) \equiv \frac{(AA) + (BB)}{\hat{k}^2} \ ,$$

$$D'_2 \equiv D'_2(A,B,k) \equiv \frac{(AA)(BB) - (AB)^2}{\hat{k}^4} \ . \quad (52)$$

The one-loop expression in eq. (41) for the effective potential in the Landau gauge can be evaluated numerically for any choice of the couplings and invariant scalar field variables $(AA)$, $(BB)$ and $(AB)^2$. As discussed before, for the study of the phase structure at weak bare couplings a good approximation can be expected, if one replaces the bare masses in the propagators by the renormalized ones. As a first approximation, one can further simplify the analysis by setting the propagator masses to zero: $m_A = m_B = m_\psi = 0$. (Later on this can be refined, for instance by combining numerical information on the renormalized masses with these perturbative formulas.) The numerical work can still become voluminous even in this simplified case, if large lattices are considered, which is important in order to diminish finite volume effects in the position and nature of the phase transitions.

Table 1: The parameters of the points where the numerical study of the one-loop effective potential was performed.

| Label | $g^2$ | $r_0$ | $r_{(AB)}$ | $\lambda_A$ | $\lambda_{(AB)}$ |
|---|---|---|---|---|---|
| a | 0.5 | 0.02 | 0.44 | 0.01 | 0.88 |
| b | 0.5 | $\sqrt{105}/210$ | $\sqrt{105}/30$ | 0.02440.. | 0.1708.. |
| c | 0.4 | $\sqrt{105}/210$ | $\sqrt{105}/30$ | 0.01952.. | 0.1366.. |
| d | 0.2 | 0.02 | 0.44 | 0.004 | 0.088 |
| e | 0.2 | 0.05 | 0.35 | 0.01 | 0.07 |

The numerical calculations were performed on $32^4$ and $64^4$ lattices, after the initial search on smaller lattices. The coupling ratios were always chosen to satisfy the relations in (2) and (4). The gauge coupling was in the range $8 \leq \beta \leq 20$ (that is $0.5 \geq g^2 \geq 0.2$), in order to stay within the weak coupling region. The parameters of five points, where most of the calculations were performed, are collected in table 1. The values of the scalar-mass parameters $m^2_{A,B}$ in the tree level potential

$$V_{\rm eff}^{\rm tree} \equiv \frac{m_A^2}{2}(AA) + \frac{m_B^2}{2}(BB) + \lambda_A(AA)^2 + \lambda_B(BB)^2 + \lambda_{[AB]}(AA)(BB) - \lambda_{(AB)}(AB)^2 \quad (53)$$

were changed as long as the phase structure was sufficiently clarified. Of course, the bare mass squares $m_A^2, m_B^2$ can also be given in terms of the scalar hopping parameters $\kappa_A, \kappa_B$, which are usually preferred in numerical simulations. The relations are:

$$\kappa_{A,B} = (8 + m^2_{A,B})^{-1} \ . \quad (54)$$

In the case where $m_\psi$ is set to zero in the fermion one-loop contribution, the effective potential does not depend on the fermion hopping parameter $\kappa_\psi$. Of course, the result is a good approximation only for such values of $\kappa_\psi$ where the renormalized fermion mass in lattice units is small. (For a perturbative estimate of the critical fermion hopping parameter, where the renormalized fermion mass vanishes, see the next section.)

For the points specified in table 1 the values of $V_{\rm eff}^{\rm tree} + V_{\rm eff}^{\rm 1-loop}$ were tabulated in an appropriate range of field variables, in order to find the minimum. In the minimal point the values of the first and second derivatives were also calculated from the analytic expressions, in order to correct for the minimum position and get an estimate of the scalar masses. In fact, choosing a fine enough mesh of points for the tabulation, the correction on the minimum position could be kept very small. The obtained phase structure in the $(\kappa_A, \kappa_B)$-plane is shown in figs. 2–6. In these figures $S$ denotes the symmetric phase with $v_A = v_B = 0$ and $A$, $B$, $AB$ the three Higgs phases, respectively.

The dangerous possibility of the appearance of first-order phase transitions is realized at the parameter values in fig. 2. In this case the radiative corrections are destroying the classical picture completely. Large parts of the renormalized parameter space are cut out by the metastable regions near the first-order transition. Fortunately, in the other cases no first-order phase transition is seen. Within the precision of the numerical calculation, no sign of a discontinuity is seen. All phase transitions are either of second order or at most very weakly first order. (Of course, a weak enough discontinuity can never be excluded by a numerical calculation with finite precision.) The only qualitative change in figs. 3–6 compared to the tree-level potential is the opening up of the degenerate line containing the degenerate $AB$-phase. (Remember that the coupling relation in (31) is satisfied due to eq. (4), therefore at tree level the $AB$-phase is degenerate at all these points.) The conclusion from these figures is that, in order to avoid a first-order phase transition, for a given value of the bare gauge coupling $g^2$ the value of $r_0$ cannot be smaller than some critical value. Otherwise the fluctuations in the lengths of the scalar fields become too strong and a first-order phase transition is induced. In the supersymmetric continuum limit, however, $r_0 \to 0$ must be realized. This requires that the critical value of $r_0$ has to become smaller and smaller for $g^2 \to 0$. Indeed, comparing figs. 2 and 5 shows that the value $r_0 = 0.02$ is well below the critical value for $g^2 = 0.5$, but comfortably above it at $g^2 = 0.2$. In order to reach a better closing of the $AB$-phase towards degeneracy at $g^2 = 0.2$, one has to go up to $r_0 = 0.05$, which is already slightly above the infrared fixed point $Q$ in fig. 1. For reaching the flow line connecting the fixed points $Q$ and $S$ with a thin wedge for the $AB$-phase one has to go to an even smaller $g^2$. The required tendency of closing the $AB$-phase towards degeneracy for fixed $(r_0, r_{(AB)})$ and decreasing $g^2$ can be seen at $Q$ by comparing fig. 3 with fig. 4. This is well displayed by the insets of these figures, showing the behaviour near the critical point $\kappa^c$ where all four phases meet.

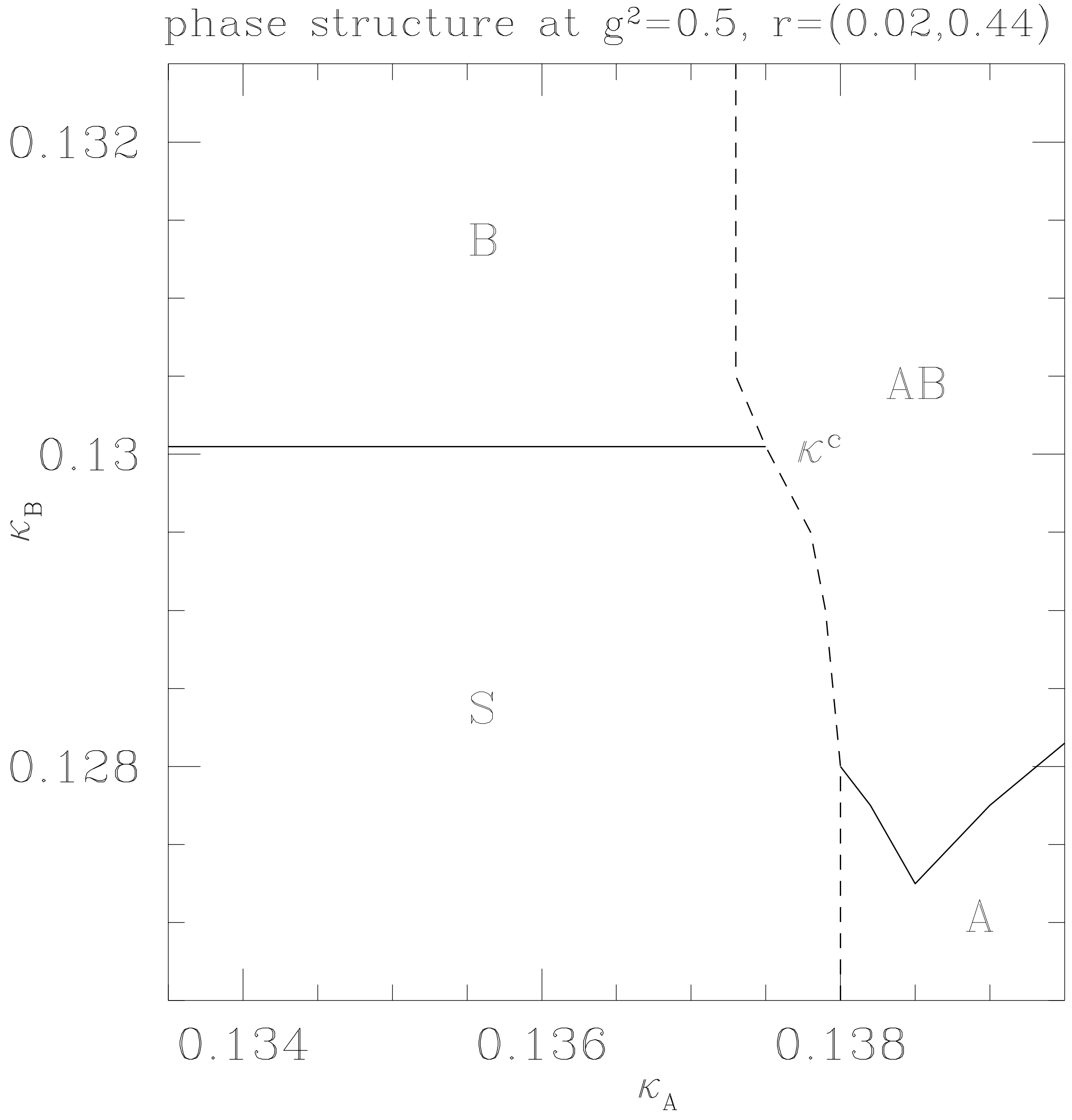

Figure 2: The phase structure in the plane of scalar hopping parameters $(\kappa_A, \kappa_B)$ at $g^2 = 0.5$, $r_0 = 0.02$, $r_{(AB)} = 0.44$. The dashed line denotes a first-order phase transition, the full lines second-order ones.

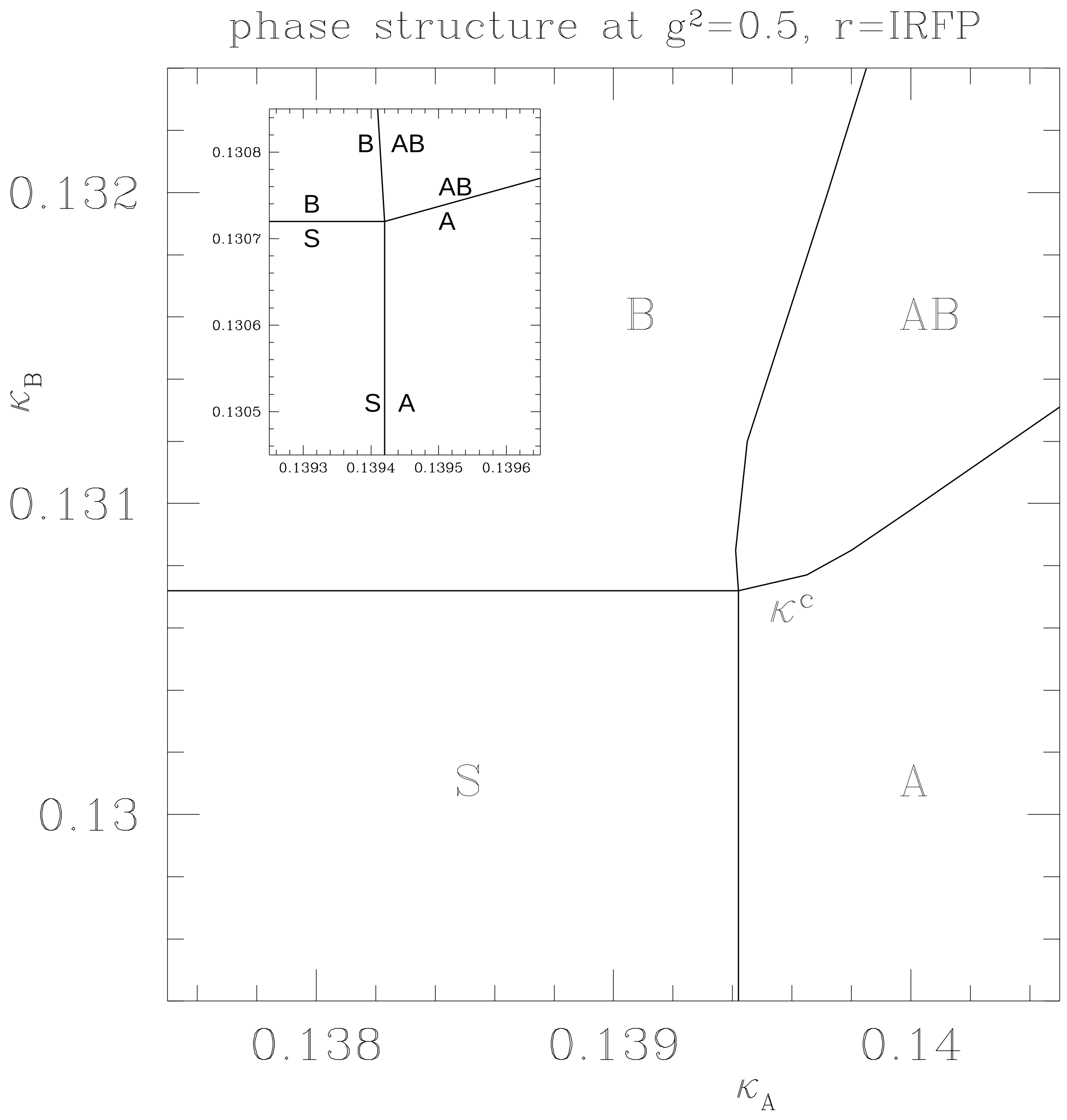

Figure 3: The phase structure in the plane of scalar hopping parameters $(\kappa_A, \kappa_B)$ at $g^2 = 0.5$ and the infrared fixed point $Q$ for the coupling ratios. The full lines denote second-order phase transitions.

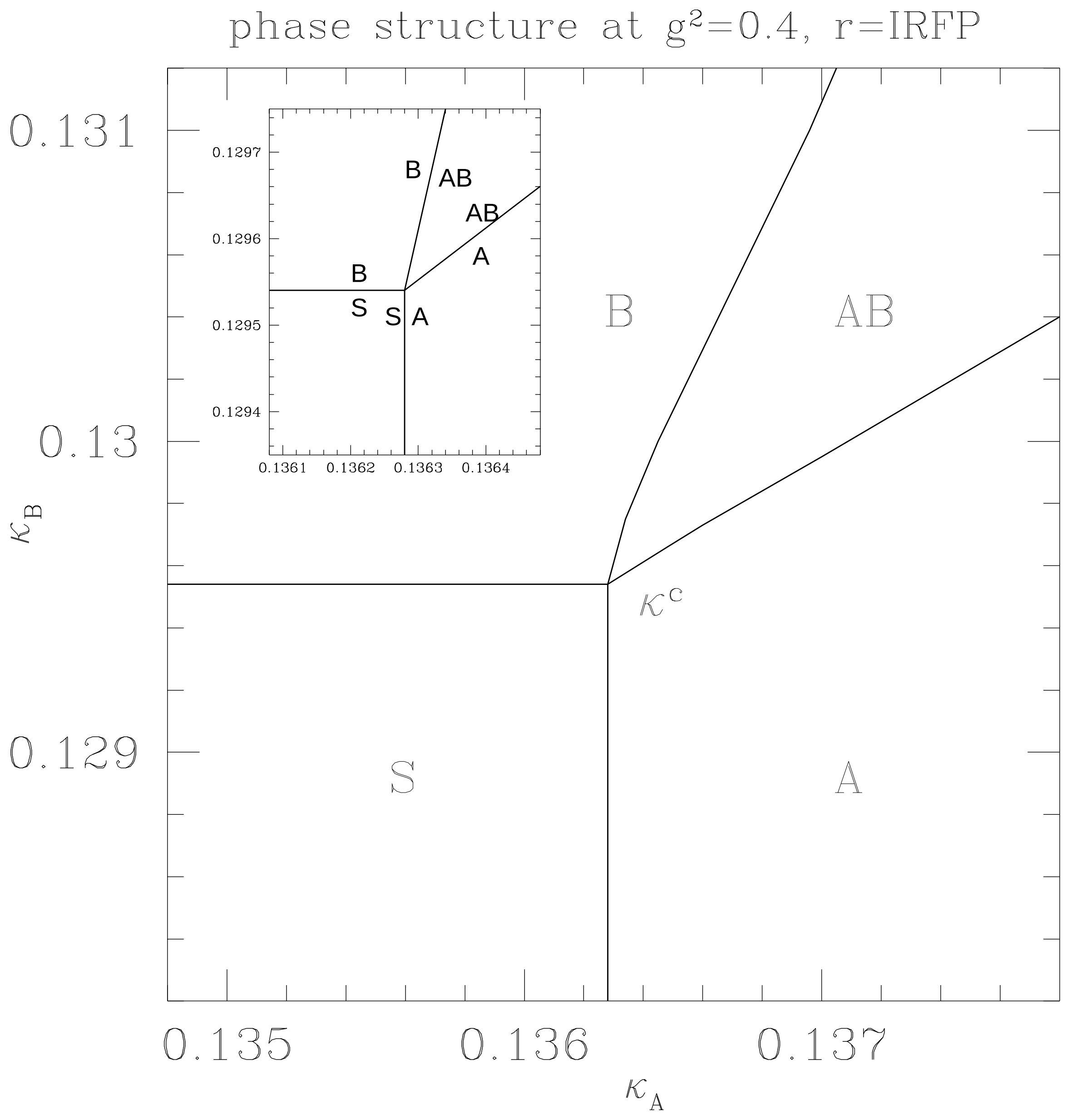

Figure 4: The phase structure in the plane of scalar hopping parameters $(\kappa_A, \kappa_B)$ at $g^2 = 0.4$ and the infrared fixed point $Q$ for the coupling ratios. The full lines denote second-order phase transitions.

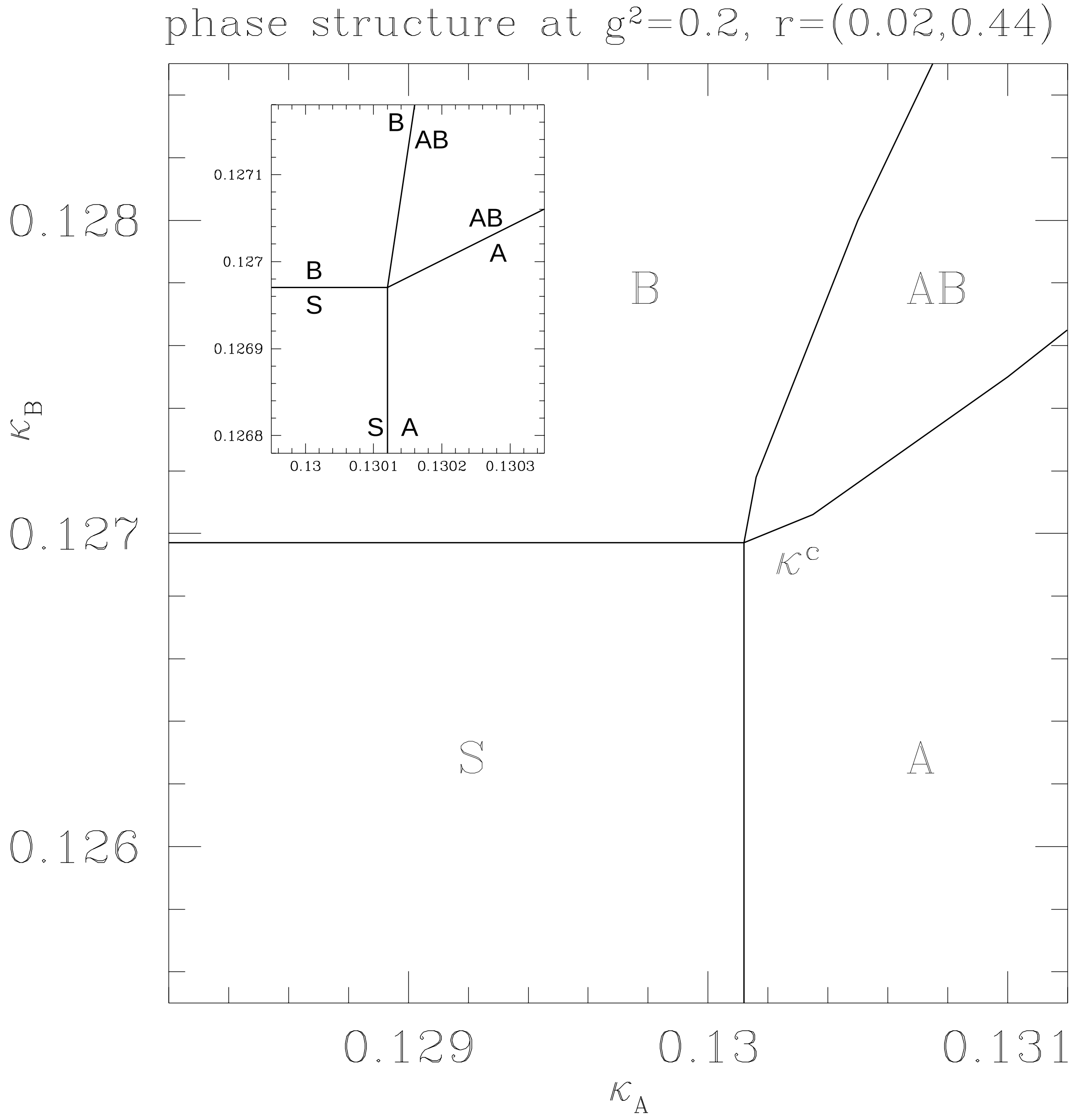

Figure 5: The phase structure in the plane of scalar hopping parameters $(\kappa_A, \kappa_B)$ at $g^2 = 0.2$, $r_0 = 0.02$, $r_{(AB)} = 0.44$. The full lines denote second-order phase transitions.

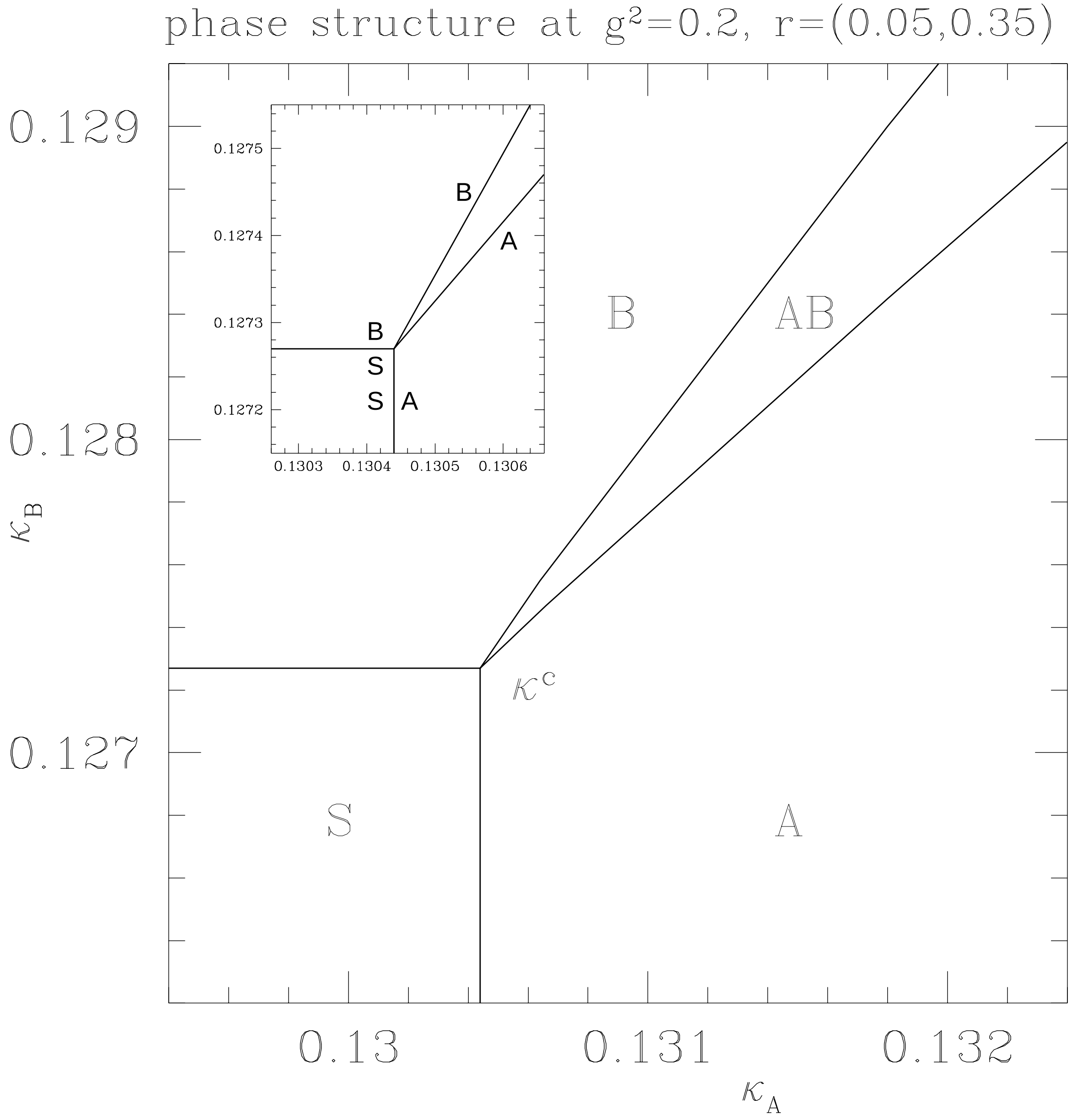

Figure 6: The phase structure in the plane of scalar hopping parameters $(\kappa_A, \kappa_B)$ at $g^2 = 0.2$, $r_0 = 0.05$, $r_{(AB)} = 0.35$. The full lines denote second-order phase transitions.

From the point of view of reaching supersymmetry, the best behaviour of the phase structure is shown in fig. 6, where the $AB$-phase is already constrained to a narrow wedge. From the experience obtained with moving around in $g^2$ and $(r_0, r_{(AB)})$, it is quite clear that the $AB$-wedge can be closed even further. This has to be checked, of course, later on by numerical simulations. Nevertheless, the proof of an exact degeneracy cannot be reached by purely numerical methods.

Another information that can be obtained from lattice perturbation theory is an estimate of the renormalized masses. What is needed is the second derivative matrix $M_2$ of the effective potential with respect to the field components, together with the $Z$-factors of wave function renormalization. The latter were not yet determined. However, general experience in Higgs-Yukawa models tells us that the values of these $Z$-factors are usually not far from 1 [9]. Neglecting their deviations from 1, mass estimates can be obtained from $M_2$ alone. An interesting question concerns the masses of the photo-axion and photo-dilaton in the $AB$-wedge, which should vanish in physical units in the $N = 2$ supersymmetric continuum limit. It turns out that in situations similar to the inset of fig. 6 the ratio of the photo-dilaton mass to the masses of the two massive scalar states is in the range of about 1/2, whereas this ratio for the photo-axion is in the range 1/5 to 1/15. As a example, the obtained mass estimates along the line $\kappa_A + \kappa_B = 0.2590$ are shown in fig. 7.

Both the smallness of the masses and the absence of discontinuities at the phase transition lines in figs. 3-6 indicate second-order or very weakly first-order phase transitions. The question of the order of the phase transitions needs, however, further study by numerical simulations and by refined perturbative analysis. For instance, one could put in the propagators of the perturbative effective potential some information on the renormalized masses either from numerical data or from perturbative calculations at moderately small masses in lattice units. This would improve the ability to disentangle weak first-order phase transitions and real second-order ones. The position and nature of the phase transitions has in any case to be investigated by numerical simulations, in order to study non-perturbative effects.

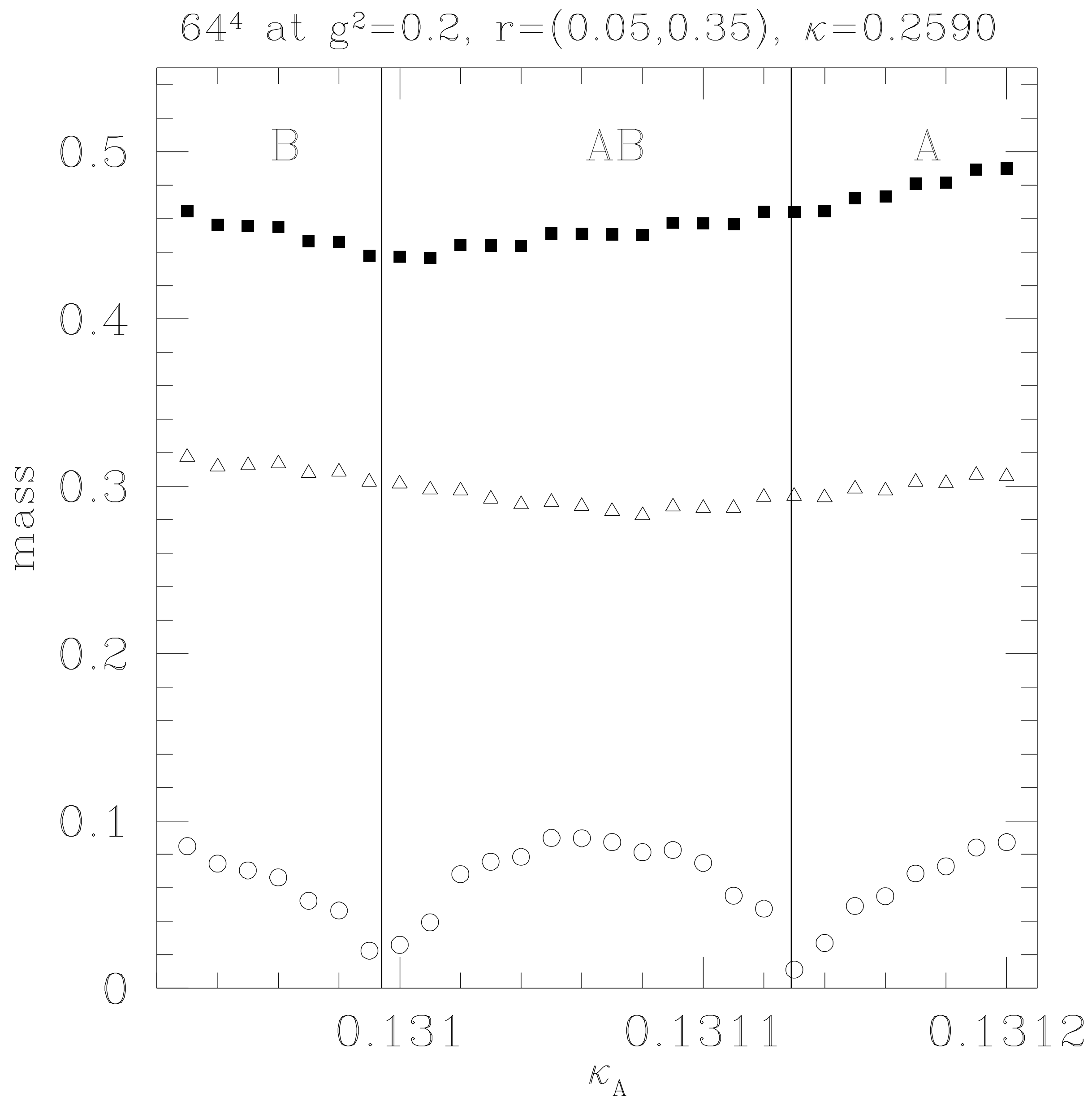

Figure 7: The mass estimates obtained from the one-loop effective potential at $g^2 = 0.2$, $r_0 = 0.05$, $r_{(AB)} = 0.35$ and $\kappa_A + \kappa_B = 0.2590$ as a function of $\kappa_A$. Full squares stand for the equal masses of the two massive scalars, open triangles for the mass of the photo-dilaton and open circles for the mass of the photo-axion. The vertical lines indicate the positions of the phase transitions.

# 5 Global symmetries

Besides the vanishing of the masses of the photo-axion and photo-dilaton, another requirement in the continuum limit is the restoration of the global $SU(2)_{\mathcal{R}}$ symmetry. In the lattice action this is broken by the fermion mass term and by the Wilson-term (proportional to $r$) to $U(1)_{\mathcal{F}}$ of fermion number conservation. The $SU(2)_{\mathcal{R}}$ symmetry can be seen in the massless continuum Euclidean action if one writes it in terms of left-handed fields only. This is achieved by replacing the right-handed components, using their relations to the charge-conjugated fields:

$$\psi_R = C\overline{\psi}_{cL}^T \ , \qquad \overline{\psi}_R = \psi_{cL}^T C \ . \tag{55}$$

Here $C$ is the usual charge-conjugation matrix for Dirac fields, which satisfies

$$C\gamma_\mu C^{-1} = -\gamma_\mu^T \ , \qquad C = -C^T = -C^{-1} \ . \tag{56}$$

The elements $U_{\mathcal{R}}$ of $SU(2)_{\mathcal{R}}$ can be defined by their action on the left-handed fields as

$$\begin{pmatrix} \psi_L \\ \psi_{cL} \end{pmatrix}' = U_{\mathcal{R}}^{-1} \begin{pmatrix} \psi_L \\ \psi_{cL} \end{pmatrix} \ , \qquad \begin{pmatrix} \overline{\psi}_L & \overline{\psi}_{cL} \end{pmatrix}' = \begin{pmatrix} \overline{\psi}_L & \overline{\psi}_{cL} \end{pmatrix} U_{\mathcal{R}} \ . \tag{57}$$

It is sometimes advantageous to introduce Majorana fields instead of the Dirac fermion field used up to now. Their definition is

$$\Psi^{(1)} \equiv \frac{1}{\sqrt{2}}(\psi + C\overline{\psi}^T) \ , \qquad \Psi^{(2)} \equiv \frac{i}{\sqrt{2}}(-\psi + C\overline{\psi}^T) \ . \tag{58}$$

These satisfy $(j = 1,2)$

$$\overline{\Psi}^{(j)} = \Psi^{(j)T} C \ . \tag{59}$$

For the left-handed fields we have

$$\begin{pmatrix} \Psi_L^{(1)} \\ \Psi_L^{(2)} \end{pmatrix} = \mathcal{S} \begin{pmatrix} \psi_L \\ \psi_{cL} \end{pmatrix} \ , \qquad \begin{pmatrix} \overline{\Psi}_L^{(1)} & \overline{\Psi}_L^{(2)} \end{pmatrix} = \begin{pmatrix} \overline{\psi}_L & \overline{\psi}_{cL} \end{pmatrix} \mathcal{S}^\dagger \ , \tag{60}$$

where the U(2) matrix $\mathcal{S}$ is given by

$$\mathcal{S} = \frac{1}{\sqrt{2}} \begin{pmatrix} 1 & 1 \\ -i & i \end{pmatrix} = \mathcal{S}^{-1\dagger} \ . \tag{61}$$

This implies that, instead of $U_{\mathcal{R}}$, the left-handed Weyl-Majorana doublet $\Psi_L^{(1,2)}$ transforms under $SU(2)_{\mathcal{R}}$ by $\mathcal{S}U_{\mathcal{R}}\mathcal{S}^\dagger$. Combining eqs. (55)–(60) one can also see that the right-handed Weyl-Majorana doublet $\Psi_R^{(1,2)}$ transforms at the same time by $\tau_2\mathcal{S}U_{\mathcal{R}}\mathcal{S}^\dagger\tau_2$, with the Pauli matrix $\tau_2$ acting on the doublet index $(j = 1,2)$.

On the Majorana basis defined by eqs. (58) and (59) the fermionic part of the lattice action is

$$S_f = \sum_x \left\{ \frac{1}{2}(m_\psi + 4r) \sum_{j=1}^{2} \overline{\Psi}_x^{(j)r} \Psi_x^{(j)r} + \epsilon_{rst} \overline{\Psi}_x^{(2)r} (G_A A_x^s + iG_B \gamma_5 B_x^s) \Psi_x^{(1)t} \right.$$

$$\left. -\frac{1}{4} \sum_{\mu=1}^{4} \sum_{j=1}^{2} \left[ \overline{\Psi}_{x+\hat{\mu}}^{(j)r} V_{rs,x\mu}(r + \gamma_\mu) \Psi_x^{(j)s} + \overline{\Psi}_x^{(j)r} V_{rs,x\mu}^{-1}(r - \gamma_\mu) \Psi_{x+\hat{\mu}}^{(j)s} \right] \right\} . \qquad (62)$$

This form is useful for the derivation of the Ward-Takahashi (WT) identities belonging to the $SU(2)_{\mathcal{R}}$ symmetry. Performing in the path integral an $x$-dependent infinitesimal $SU(2)_{\mathcal{R}}$ transformation:

$$\Psi'_x = (1 - i\alpha_{1x}\gamma_5\tau_1 + i\alpha_{2x}\tau_2 + i\alpha_{3x}\gamma_5\tau_3)\Psi_x , \qquad \overline{\Psi}'_x = \overline{\Psi}_x(1 - i\alpha_{1x}\gamma_5\tau_1 - i\alpha_{2x}\tau_2 + i\alpha_{3x}\gamma_5\tau_3) , \quad (63)$$

with infinitesimal $\alpha_{jx}$, the invariance of the path integral measure implies the WT identities.

First of all, for the unbroken $U(1)_{\mathcal{F}}$ subgroup the gauge-invariant vector current of fermion number can be defined as

$$J_{x\mu}^{\mathcal{F}} \equiv J_{x\mu}^{(2)} = \frac{1}{4} \left[ -\overline{\Psi}_{x+\hat{\mu}}\tau_2 V_{x\mu}(r + \gamma_\mu)\Psi_x + \overline{\Psi}_x \tau_2 V_{x\mu}^{-1}(r - \gamma_\mu)\Psi_{x+\hat{\mu}} \right] . \qquad (64)$$

This satisfies the unbroken WT identity

$$0 = \left\langle \sum_{\mu=1}^{4} (J_{x\mu}^{\mathcal{F}} - J_{x-\hat{\mu},\mu}^{\mathcal{F}}) \right\rangle \equiv \left\langle \Delta_\mu^b J_{x\mu}^{\mathcal{F}} \right\rangle . \qquad (65)$$

The other two gauge-invariant currents $J_{x\mu}^{(j=1,3)}$, which can be defined as

$$J_{x\mu}^{(j)} = \frac{1}{4}(-1)^{(j-1)/2} \left[ \overline{\Psi}_{x+\hat{\mu}}\tau_j\gamma_\mu\gamma_5 V_{x\mu}\Psi_x + \overline{\Psi}_x\tau_j\gamma_\mu\gamma_5 V_{x\mu}^{-1}\Psi_{x+\hat{\mu}} \right] , \qquad (66)$$

correspond to the other two generators of $SU(2)_{\mathcal{R}}$. The corresponding symmetries are explicitly broken by the fermion mass $m_\psi$ and by the Wilson term proportional to $r$; therefore the currents $J_{x\mu}^{(j=1,3)}$ satisfy the broken WT identities

$$\left\langle \Delta_\mu^b J_{x\mu}^{(j)} \right\rangle = (-1)^{(j-1)/2} \left\langle m_\psi \overline{\Psi}_x \tau_j \gamma_5 \Psi_x + X_x^{(j)} \right\rangle . \qquad (67)$$

Here the symmetry-breaking term proportional to $r$ is given by

$$X_x^{(j)} \equiv \frac{r}{4} \sum_{\mu=1}^{4} \left[ 4\overline{\Psi}_x\tau_j\gamma_5\Psi_x - \overline{\Psi}_{x+\hat{\mu}}\tau_j\gamma_5 V_{x\mu}\Psi_x - \overline{\Psi}_x\tau_j\gamma_5 V_{x\mu}^{-1}\Psi_{x+\hat{\mu}} \right.$$

$$\left. -\overline{\Psi}_{x-\hat{\mu}}\tau_j\gamma_5 V_{x-\hat{\mu},\mu}^{-1}\Psi_x - \overline{\Psi}_x\tau_j\gamma_5 V_{x-\hat{\mu},\mu}\Psi_{x-\hat{\mu}} \right] . \qquad (68)$$

The WT identity in (65) corresponds to exact fermion number conservation at non-zero lattice spacing. The broken WT identities in (67) are very similar to the flavour non-singlet axial-vector WT identities in lattice QCD with Wilson quarks [10]. Taking into account operator mixings, and appropriately redefining the current normalizations and the critical hypersurface for zero fermion mass, one-loop lattice perturbation theory suggests that in the massless continuum limit these identities are reproducing the expected unbroken WT identities in the target continuum theory.

Of course, the zero fermion-mass hypersurface defined in this way corresponds only to one possible definition at non-zero lattice spacings. There are many other definitions, which all tend, according to lattice perturbation theory, to the same hypersurface in the continuum limit. One possibility is to calculate the fermion self-energy $\Sigma_\psi$ in the $AB$-phase at vanishing four-momentum $p = 0$. Applying the Feynman rules of section 3 one obtains the general form

$$\Sigma_\psi(p=0) = M^{\psi 1}_{r_1 r_2} + i\gamma_5 M^{\psi 2}_{r_1 r_2} \; . \tag{69}$$

The matrices in isospin indices $r_1, r_2$ satisfy, at non-zero lattice spacing:

$$0 = M^{\psi 1}_{11} - M^{\psi 1}_{22} = M^{\psi 1}_{12} + M^{\psi 1}_{21} = M^{\psi 1}_{13} = M^{\psi 1}_{31} = M^{\psi 1}_{23} = M^{\psi 1}_{32} \; ,$$

$$0 = M^{\psi 5}_{11} - M^{\psi 5}_{22} = M^{\psi 5}_{12} + M^{\psi 5}_{21} = M^{\psi 5}_{13} = M^{\psi 5}_{31} = M^{\psi 5}_{23} = M^{\psi 5}_{32} \; . \tag{70}$$

In the continuum limit, when in lattice units $v_A, v_B \to 0$, the only non-zero matrix elements are

$$M^{\psi 1}_{11} = M^{\psi 1}_{33} = \frac{1}{\Omega}\sum_k \frac{4(G_A^2 - G_B^2 - 4g^2)\hat{k}^2 - 8g^2\bar{k}^2 - 2g^2\hat{k}^4}{\hat{k}^2(4\bar{k}^2 + \hat{k}^4)} \; . \tag{71}$$

Here we have set, for simplicity, the Wilson parameter to $r = 1$. The simplest definition of the critical value of the bare fermion mass $m^c_\psi$ is to require $m^c_\psi + M^{\psi 1}_{11} = 0$. This gives

$$m^c_\psi = \frac{1}{\Omega}\sum_k \left[\frac{4(4g^2 + G_B^2 - G_A^2)}{4\bar{k}^2 + \hat{k}^4} + \frac{2g^2}{\hat{k}^2}\right] = (4g^2 + G_B^2 - G_A^2)\cdot 0.085.. + g^2 \cdot 0.310.. \tag{72}$$

Remember that the critical fermion hopping parameter can be obtained from $m^c_\psi$ by $\kappa^c_\psi = 1/(8 + 2m^c_\psi)$.

Let us note that from $\Sigma_\psi$ one can also extract the one-loop corrections to the renormalized Yukawa couplings $G^{\rm ren}_{A,B}$. They turn out to be, respectively,

$$\lim_{v_A, v_B \to 0} \frac{M^{\psi 1}_{12}}{i v_A} = \frac{G_A}{\Omega}\sum_k \frac{16g^2\bar{k}^4 + 4(G_A^2 - G_B^2 - 4g^2)(4\hat{k}^2\bar{k}^2 - \hat{k}^6) - g^2\hat{k}^8}{\hat{k}^4(4\bar{k}^2 + \hat{k}^4)^2} \; ,$$

$$\lim_{v_A, v_B \to 0} -\frac{M^{\psi 5}_{12}}{v_B} = \frac{G_B}{\Omega}\sum_k \frac{g^2\hat{k}^4 + 4(G_B^2 - G_A^2 - 4g^2)\hat{k}^2 + 4g^2\bar{k}^2}{\hat{k}^4(4\bar{k}^2 + \hat{k}^4)} \; . \tag{73}$$

The other important global symmetry besides $\mathrm{SU}(2)_{\mathcal{R}}$, which has to be restored in the continuum limit is, of course, $N = 2$ supersymmetry. In order to discuss the supersymmetry transformations, it is advantageous to introduce the so-called "symplectic Majorana" fields [11], which transform simply as a doublet under $\mathrm{SU}(2)_{\mathcal{R}}$. These are related to the Majorana fields in (58) by

$$\lambda \equiv (\tau_2 P_R + P_L)\Psi \; , \qquad \overline{\lambda} \equiv \overline{\Psi}(\tau_2 P_L + P_R) \; . \tag{74}$$

Here $P_R = (1+\gamma_5)/2$, $P_L = (1-\gamma_5)/2$ denote the chiral projectors and the Pauli matrix $\tau_2$ acts on the $\mathrm{SU}(2)_{\mathcal{R}}$ doublet indices $j = 1,2$. Instead of eq. (59), these satisfy the symplectic Majorana relations

$$\overline{\lambda} = -\lambda^T C\gamma_5\tau_2 \ , \qquad \lambda = C\gamma_5\tau_2\overline{\lambda}^T \ . \tag{75}$$

The $N = 2$ supersymmetric Euclidean action in the continuum can be written with these fields as

$$S_{\rm susy} = \int d^4x \Big\{ \frac{1}{4}F^r_{\mu\nu}(x)F^r_{\mu\nu}(x) + \frac{1}{2}(D_\mu A^r(x))(D_\mu A^r(x)) + \frac{1}{2}(D_\mu B^r(x))(D_\mu B^r(x))$$
$$+\frac{1}{2}\overline{\lambda}^{jr}(x)\gamma_\mu D_\mu \lambda^{jr}(x) - \frac{ig}{2}\epsilon_{rst}\overline{\lambda}^{jr}(x)[A^s(x) + i\gamma_5 B^s(x)]\lambda^{jt}(x)$$
$$+\frac{g^2}{2}[A^r(x)A^r(x)B^s(x)B^s(x) - A^r(x)B^r(x)A^s(x)B^s(x)] \Big\} \ , \tag{76}$$

where the notations for the field strength and the covariant derivatives are, as usual,

$$F^r_{\mu\nu} \equiv \partial_\mu A^r_\nu(x) - \partial_\nu A^r_\mu(x) + g\epsilon_{rst}A^s_\mu(x)A^t_\nu(x) \ , \quad D_\mu A^r(x) \equiv \partial_\mu A^r(x) + g\epsilon_{rst}A^s_\mu(x)A^t(x) \ . \tag{77}$$

Global $N = 2$ supersymmetry is realized non-linearly in a formulation without auxiliary fields [11]. Introducing the symplectic Majorana doublet Grassmann variable $\epsilon$, which satisfies the same relations as (75)

$$\overline{\epsilon} = -\epsilon^T C\gamma_5\tau_2 \ , \qquad \epsilon = C\gamma_5\tau_2\overline{\epsilon}^T \ , \tag{78}$$

the supersymmetry transformations are defined as

$$\delta A^r = \overline{\epsilon}\lambda^r \ , \qquad \delta B^r = i\overline{\epsilon}\gamma_5\lambda^r \ , \qquad \delta A^r_\mu = i\overline{\epsilon}\gamma_\mu\lambda^r \ ,$$
$$\delta\lambda^r = -\frac{1}{2}F^r_{\mu\nu}\sigma_{\mu\nu}\epsilon - \gamma_\mu(D_\mu A^r + i\gamma_5 D_\mu B^r)\epsilon + g\epsilon_{rst}A^s B^t \gamma_5\epsilon \ . \tag{79}$$

Here $\sigma_{\mu\nu} \equiv (i/2)[\gamma_\mu, \gamma_\nu]$ is used. By partial integration and neglecting surface terms, it can be shown that (79) is a symmetry of the above action:

$$\delta S_{\rm susy} = 0 \ . \tag{80}$$

There is a gauge-invariant conserved current corresponding to this symmetry. It is the $\mathrm{SU}(2)_{\mathcal{R}}$ doublet spinor-vector current

$$J^{\rm susy}_\mu(x) \equiv \Big\{ -\frac{1}{4}F^r_{\rho\sigma}(x)\sigma_{\rho\sigma} + \frac{1}{2}D_\rho[A^r(x) + i\gamma_5 B^r(x)]\gamma_\rho - \frac{g}{2}\epsilon_{rst}A^s(x)B^t(x)\gamma_5 \Big\} \gamma_\mu\lambda^r(x) \ . \tag{81}$$

As a consequence of the field equations of motion, the divergence of $J^{\rm susy}_\mu$ vanishes.

For the study of the supersymmetric WT identities, one has to define the transformations (79) on the lattice. This requires some definition of the field strength tensor $F^{\rm lattice}_{x\mu\nu}$, for instance the usual "clover" definition averaging over the four open plaquettes in the $\mu,\nu$-plane, which are touching the site $x$. Since the gauge link variables are given by $U_{x\mu} = \exp(igA^r_{x\mu}T_r)$, the transformation of $A^r_{x\mu}$ can be represented, for instance, by

$$\delta U_{x\mu} = -g\overline{\epsilon}\gamma_\mu\lambda^r T_r U_{x\mu} \ . \tag{82}$$

By eq. (16) this also implies the transformation of the gauge links in the adjoint representation $V_{rs,x\mu}$. The covariant derivatives on the lattice are replaced by covariant differences defined, for instance, as

$$\Delta(U)_\mu A_x^r \equiv \frac{1}{2}\left[V_{rs,x\mu}^{-1} A_{x+\hat{\mu}}^s - V_{rs,x-\hat{\mu},\mu} A_{x-\hat{\mu}}^s\right] \ . \tag{83}$$

All these choices contain, of course, some arbitrariness at the level of $\mathcal{O}(a)$ corrections, which go to zero by the lattice spacing $a$.

Having defined the supersymmetry transformations on the lattice, the broken WT identities corresponding to $N = 2$ supersymmetry can be derived. Due to the involved operator mixings, the analysis of the continuum limit is in this case more subtle than in the $N = 1$ case discussed in ref. [3] and will not be considered in the present paper.

# 6 Discussion and conclusion

The flow pattern of the renormalization group trajectories at weak couplings in the SU(2) adjoint Higgs-Yukawa model makes it possible to tune the coupling ratios to the values required by $N = 2$ supersymmetry. The fixed-point structure is particularly transparent in the simplified version of the renormalization group equations studied in section 2, where several coupling ratios are set to satisfy relations that are left invariant by the complete set of one-loop equations in ref. [5]. It is interesting that, besides the supersymmetric fixed point of coupling ratios $S$, there is also another fixed point $Q$, which is attractive in the infrared limit (see fig. 1). This fixed point is particularly important for the behaviour of the ratios of renormalized couplings, if the bare couplings are kept fixed, because the flow directions are then reversed. In order to avoid the attraction of this non-supersymmetric fixed point, one has to reach the separating flow line that leads, in the infrared limit, to the fixed point $S$ from the direction $\mathbf{e}_2$ in eq. (11). However, it should be kept in mind that in the infrared limit the couplings become strong, and therefore the unknown higher-order corrections become important.

The $N = 2$ supersymmetric Yang-Mills theory has a weak coupling limit, when the expectation values of the gauge-scalar fields get large compared with the $\Lambda$-parameter of the asymptotically free gauge coupling. This is a great advantage, because in this region the theory is perturbative. In a lattice formulation one can use the methods of lattice perturbation theory, which can support the numerical simulations and contain a lot of important information concerning the phase structure, masses, renormalized couplings, etc.

A first look at the phase structure, according to the one-loop lattice effective potential in the Landau gauge, shows several qualitative features which are very important from the point of view of the existence of the supersymmetric continuum limit:

- If the quartic couplings $\lambda_{A,B}$ stabilizing the path integral are not too small, then the phase transitions are of second order or very weakly first order. This is the prerequisite for the possibility of a continuum limit.
- The critical value of the ratio $r_0 = \lambda_{A,B}/g^2$, above which the first-order phase transition spoiling the continuum limit disappears, gets rapidly smaller by decreasing $g^2$. This allows us, for small enough values of $g^2$, to reach the flow line of renormalization group trajectories, which leads to the supersymmetric continuum limit.

- The wedge containing the $AB$-phase can be tuned to become narrow, presumably even infinitely narrow in the continuum limit. This gives zero mass for the photo-axion, corresponding to a flat direction in quantum moduli space. The small mass of the photo-axion is also supported by the mass estimates from the effective potential (see fig. 7).

- The broken WT identities corresponding to the global $SU(2)_{\mathcal{R}}$ symmetry are very similar to the WT identities for flavour non-singlet axial symmetry in QCD with Wilson fermions. Therefore, it is plausible that the $SU(2)_{\mathcal{R}}$ symmetry is restored in the continuum limit if the renormalized fermion mass is tuned to zero.

These constitute all but one conditions for the existence of the supersymmetric continuum limit. The open question is the mass of the photo-dilaton, where the present investigation is inconclusive. The left-over combination of the three bare mass parameters, after tuning to a zero fermion mass and infinitely narrow $AB$-wedge, could be a relevant mass parameter for the overall length of the vacuum expectation values denoted by $v/\Lambda \equiv (v_A^2 + v_B^2)^{\frac{1}{2}}/\Lambda$. Then in the continuum limit the mass of the photo-dilaton would be a function of $v/\Lambda$, which is determined by the dynamics and does not need to be identically zero. (In case the width of the $AB$-wedge cannot be tuned to exactly zero in the continuum limit, the same would also hold for the mass of the photo-axion.) This expectation seems to be strengthened also by the presence of a first-order phase transition at small $\lambda_{A,B}$, which can imply a lower limit for the mass of the photo-dilaton. Nevertheless, it is very possible that the first-order phase transition and the relatively large mass estimate obtained for the photo-dilaton can be avoided by better parameter tuning. For instance, one could exploit the above-mentioned infrared attractive direction $\mathbf{e}_2$ in eq. (11), which leads to supersymmetric coupling ratios for renormalized couplings and hence to a massless photo-dilaton. This question deserves further investigation.

Further information on the supersymmetric continuum limit can be obtained by numerical simulations and by a refinement of the perturbative analysis. After the question of the continuum limit has been sufficiently clarified, the predictions of ref. [1] could be exploited in this lattice realization of SYM2. The behaviour of the phase transition near the critical point where all phases, including the symmetric phase, meet could explain, in the supersymmetric low energy effective theory, the absence of a singularity corresponding to the symmetric phase with massless elementary fields. The interesting predictions concerning the spectrum of light states, including massless monopoles, could be investigated. Since the lattice action contains the pure SU(2) Yang-Mills theory as a special case, the understanding of the confinement mechanism in the supersymmetric part of the parameter space could be transferred to pure gauge theory.

## Acknowledgements

I thank Gabriele Veneziano, Zoltán Fodor, Rainer Sommer and Shimon Yankielowicz for helpful discussions, and Karl Jansen for correspondence on the effective potential.

# A Appendix: the contribution of the scalar loops

In this appendix the contribution of the scalar loops to the one-loop effective potential is given in the special case where

$$\lambda_B = \lambda_A \ , \qquad \lambda_{[AB]} = \lambda_{(AB)} + 2\lambda_A \ . \tag{84}$$

The notations are somewhat different from those in section 4, namely:

$$\lambda_{ABAB} \equiv \lambda_{(AB)} \ , \qquad AA \equiv (AA) \ , \qquad BB \equiv (BB) \ , \qquad AB \equiv (AB) \ . \tag{85}$$

The expression is ordered according to the powers of

$$D_A \equiv (m_A^2 + \hat{k}^2)^{-1} \ , \qquad D_B \equiv (m_B^2 + \hat{k}^2)^{-1} \tag{86}$$

and $\lambda_{ABAB}, \lambda_A$. Monomials of these variables are followed in the next one or two lines by their coefficients in $\mathcal{D}_s$ required in eq. (42):

$$1$$

$$1$$

$$D_A \lambda_{ABAB}$$

$$4\, BB$$

$$D_A \lambda_A$$

$$20\, AA + 12\, BB$$

$$D_B \lambda_{ABAB}$$

$$4\, AA$$

$$D_B \lambda_A$$

$$12\, AA + 20\, BB$$

$${D_A}^2 {\lambda_{ABAB}}^2$$

$$4\, BB^2$$

$${D_A}^2 \lambda_A \lambda_{ABAB}$$

$$48\, AA\, BB + 32\, BB^2 + 16\, AB^2$$

$${D_A}^2 {\lambda_A}^2$$

$$16\, (AA + BB)\, (7\, AA + 3\, BB)$$

$$D_B D_A {\lambda_{ABAB}}^2$$

$$12\, AB^2 - 4\, AA\, BB$$

$$D_B D_A \lambda_A \lambda_{ABAB}$$

$$80\, AA^2 + 64\, AB^2 + 32\, AA\, BB + 80\, BB^2$$

$$D_B D_A {\lambda_A}^2$$

$$240\ (AA + BB)^2$$

$${D_B}^2 {\lambda_{ABAB}}^2$$

$$4\,AA^2$$

$${D_B}^2 \lambda_A \lambda_{ABAB}$$

$$32\,AA^2 + 16\,AB^2 + 48\,AA\,BB$$

$${D_B}^2 {\lambda_A}^2$$

$$16\ (3\,AA + 7\,BB)\,(AA + BB)$$

$${D_A}^3 \lambda_A {\lambda_{ABAB}}^2$$

$$16\,BB\ \left(BB^2 + AA\,BB + 2\,AB^2\right)$$

$${D_A}^3 {\lambda_A}^2 \lambda_{ABAB}$$

$$64\ (AA + BB)\left(2\,AA\,BB + BB^2 + AB^2\right)$$

$${D_A}^3 {\lambda_A}^3$$

$$64\ (3\,AA + BB)\,(AA + BB)^2$$

$$D_B {D_A}^2 {\lambda_{ABAB}}^3$$

$$-32\,BB\ \left(AA\,BB - AB^2\right)$$

$$D_B {D_A}^2 \lambda_A {\lambda_{ABAB}}^2$$

$$80\,BB^3 + 128\,AA\,AB^2 - 112\,AA\,BB^2 + 224\,BB\,AB^2$$

$$D_B {D_A}^2 {\lambda_A}^2 \lambda_{ABAB}$$

$$64\ (AA + BB)\left(7\,AA^2 + 4\,AA\,BB + 11\,AB^2 + 10\,BB^2\right)$$

$$D_B {D_A}^2 {\lambda_A}^3$$

$$192\ (7\,AA + 5\,BB)\,(AA + BB)^2$$

$${D_B}^2 D_A {\lambda_{ABAB}}^3$$

$$-32\,AA\ \left(AA\,BB - AB^2\right)$$

$${D_B}^2 D_A \lambda_A {\lambda_{ABAB}}^2$$

$$224\,AA\,AB^2 + 128\,BB\,AB^2 - 112\,AA^2 BB + 80\,AA^3$$

$${D_B}^2 D_A {\lambda_A}^2 \lambda_{ABAB}$$

$$64\ (AA + BB)\left(10\,AA^2 + 4\,AA\,BB + 11\,AB^2 + 7\,BB^2\right)$$

$${D_B}^2 D_A {\lambda_A}^3$$

$$192\ (5\,AA + 7\,BB)\,(AA + BB)^2$$

$${D_B}^3 \lambda_A {\lambda_{ABAB}}^2$$

$$16\,AA\ \left(AA^2 + AA\,BB + 2\,AB^2\right)$$

$$D_B{}^3\lambda_A{}^2\lambda_{ABAB}$$

$$64\ (AA+BB)\left(AA^2+2\,AA\,BB+AB^2\right)$$

$$D_B{}^3\lambda_A{}^3$$

$$64\ (AA+3\,BB)\,(AA+BB)^2$$

$$D_B D_A{}^3\lambda_{ABAB}{}^4$$

$$-16\,BB^2\left(AA\,BB-AB^2\right)$$

$$D_B D_A{}^3\lambda_A\lambda_{ABAB}{}^3$$

$$-64\ \left(3\,AA\,BB+2\,BB^2-AB^2\right)\left(AA\,BB-AB^2\right)$$

$$D_B D_A{}^3\lambda_A{}^2\lambda_{ABAB}{}^2$$

$$64\ (AA+BB)\left(AA^2BB-6\,AA\,BB^2+5\,AA\,AB^2+17\,BB\,AB^2+5\,BB^3\right)$$

$$D_B D_A{}^3\lambda_A{}^3\lambda_{ABAB}$$

$$256\ (AA+BB)^2\left(3\,AA^2+3\,AA\,BB+5\,BB^2+7\,AB^2\right)$$

$$D_B D_A{}^3\lambda_A{}^4$$

$$256\ (9\,AA+5\,BB)\,(AA+BB)^3$$

$$D_B{}^2D_A{}^2\lambda_{ABAB}{}^4$$

$$-16\ \left(AA\,BB-AB^2\right)\left(3\,AA\,BB-AB^2\right)$$

$$D_B{}^2D_A{}^2\lambda_A\lambda_{ABAB}{}^3$$

$$-64\ \left(AA\,BB-AB^2\right)\left(5\,AA^2+8\,AA\,BB+5\,BB^2+2\,AB^2\right)$$

$$D_B{}^2D_A{}^2\lambda_A{}^2\lambda_{ABAB}{}^2$$

$$64\ (AA+BB)^2\left(7\,AA^2-20\,AA\,BB+7\,BB^2+32\,AB^2\right)$$

$$D_B{}^2D_A{}^2\lambda_A{}^3\lambda_{ABAB}$$

$$512\ (AA+BB)^2\left(7\,AA^2+AA\,BB+7\,BB^2+11\,AB^2\right)$$

$$D_B{}^2D_A{}^2\lambda_A{}^4$$

$$5376\ (AA+BB)^4$$

$$D_B{}^3D_A\lambda_{ABAB}{}^4$$

$$-16\,AA^2\left(AA\,BB-AB^2\right)$$

$$D_B{}^3D_A\lambda_A\lambda_{ABAB}{}^3$$

$$-64\ \left(AA\,BB-AB^2\right)\left(2\,AA^2+3\,AA\,BB-AB^2\right)$$

$$D_B{}^3D_A\lambda_{ABAB}{}^2\lambda_A{}^2$$

$$64\ (AA+BB)\left(5\,AA^3-6\,AA^2BB+AA\,BB^2+17\,AA\,AB^2+5\,BB\,AB^2\right)$$

$$D_B{}^3D_A\lambda_{ABAB}\lambda_A{}^3$$

$$256\ (AA+BB)^2\left(5\,AA^2+3\,AA\,BB+7\,AB^2+3\,BB^2\right)$$

$$D_B{}^3D_A\lambda_A{}^4$$

$$256\ (5\,AA+9\,BB)\,(AA+BB)^3$$

$$D_B{}^2D_A{}^3\lambda_{ABAB}{}^5$$

$$64\,BB\ \left(AA\,BB-AB^2\right)^2$$

$$D_B{}^2D_A{}^3\lambda_A\lambda_{ABAB}{}^4$$

$$-64\ \left(AA\,BB-AB^2\right)$$

$$\cdot\left(7\,AA^2BB-3\,AA\,BB^2-3\,AA\,AB^2+4\,BB^3+11\,BB\,AB^2\right)$$

$$D_B{}^2D_A{}^3\lambda_A{}^2\lambda_{ABAB}{}^3$$

$$-256\ (3\,AA+5\,BB)\,(AA+BB)^2\left(AA\,BB-AB^2\right)$$

$$D_B{}^2D_A{}^3\lambda_A{}^3\lambda_{ABAB}{}^2$$

$$256\ (AA+BB)^2$$

$$\cdot\left(3\,AA^3-5\,AA^2BB-13\,AA\,BB^2+18\,AA\,AB^2+30\,BB\,AB^2+7\,BB^3\right)$$

$$D_B{}^2D_A{}^3\lambda_A{}^4\lambda_{ABAB}$$

$$1024\ (AA+BB)^3\left(6\,AA^2+AA\,BB+7\,BB^2+12\,AB^2\right)$$

$$D_B{}^2D_A{}^3\lambda_A{}^5$$

$$1024\ (9\,AA+7\,BB)\,(AA+BB)^4$$

$$D_B{}^3D_A{}^2\lambda_{ABAB}{}^5$$

$$64\,AA\ \left(AA\,BB-AB^2\right)^2$$

$$D_B{}^3D_A{}^2\lambda_A\lambda_{ABAB}{}^4$$

$$-64\ \left(AA\,BB-AB^2\right)$$

$$\cdot\left(4\,AA^3-3\,AA^2BB+7\,AA\,BB^2+11\,AA\,AB^2-3\,BB\,AB^2\right)$$

$$D_B{}^3D_A{}^2\lambda_A{}^2\lambda_{ABAB}{}^3$$

$$-256\ (5\,AA+3\,BB)\,(AA+BB)^2\left(AA\,BB-AB^2\right)$$

$$D_B{}^3D_A{}^2\lambda_A{}^3\lambda_{ABAB}{}^2$$

$$256\ (AA+BB)^2$$

$$\cdot\left(7\,AA^3-13\,AA^2BB-5\,AA\,BB^2+30\,AA\,AB^2+18\,BB\,AB^2+3\,BB^3\right)$$

$$D_B{}^3D_A{}^2\lambda_A{}^4\lambda_{ABAB}$$

$$1024\ (AA+BB)^3\left(7\,AA^2+AA\,BB+6\,BB^2+12\,AB^2\right)$$

$$D_B{}^3D_A{}^2\lambda_A{}^5$$

$$1024\ (7\,AA + 9\,BB)\,(AA + BB)^4$$

$${D_B}^3{D_A}^3{\lambda_{ABAB}}^6$$

$$192\ \left(AA\,BB - AB^2\right)^3$$

$${D_B}^3{D_A}^3\lambda_A{\lambda_{ABAB}}^5$$

$$1536\ \left(AA\,BB - AB^2\right)^3$$

$${D_B}^3{D_A}^3{\lambda_A}^2{\lambda_{ABAB}}^4$$

$$-768\ (AA + BB)^2\left(AA\,BB - AB^2\right)\left(AA^2 - AA\,BB + BB^2 + 3\,AB^2\right)$$

$${D_B}^3{D_A}^3{\lambda_A}^3{\lambda_{ABAB}}^3$$

$$-3072\ (AA + BB)^4\left(AA\,BB - AB^2\right)$$

$${D_B}^3{D_A}^3{\lambda_A}^4{\lambda_{ABAB}}^2$$

$$3072\ (AA + BB)^4\left(AA^2 - 3\,AA\,BB + BB^2 + 5\,AB^2\right)$$

$${D_B}^3{D_A}^3{\lambda_A}^5\lambda_{ABAB}$$

$$12288\ (AA + BB)^4\left(AA^2 + BB^2 + 2\,AB^2\right)$$

$${D_B}^3{D_A}^3{\lambda_A}^6$$

$$12288\ (AA + BB)^6$$